\documentclass[prl,superscriptaddress,twocolumn,notitlepage,longbibliography]{revtex4-2}
\usepackage[unicode=true, bookmarks=true,bookmarksnumbered=false,bookmarksopen=false,
 breaklinks=false,pdfborder={0 0 1},backref=false,colorlinks=true]
 {hyperref}
\hypersetup{
linkcolor=magenta,urlcolor=blue,citecolor=blue,pdfstartview={FitH},urlcolor=blue}

\usepackage{amsfonts}
\usepackage{subfigure}
\usepackage{amsmath}
\usepackage{amssymb}
\usepackage{amsbsy} 
\usepackage{epsfig}
\usepackage{graphicx}
\usepackage{epstopdf}
\usepackage{bbm}
\usepackage{xcolor}
\usepackage{minitoc}

\begin{document}

\title{Non-Hermitian Delocalization Realizes Random Dirac Criticality in One Dimension}

\author{Bo Li}
\email{libphysics@xjtu.edu.cn}
\affiliation{MOE Key Laboratory for Nonequilibrium Synthesis and Modulation of Condensed Matter,\\
Shaanxi Province Key Laboratory of Quantum Information and Quantum Optoelectronic Devices,\\
School of Physics, Xi’an Jiaotong University, Xi’an 710049, China}

\author{Shen Zhang}
\affiliation{MOE Key Laboratory for Nonequilibrium Synthesis and Modulation of Condensed Matter,\\
Shaanxi Province Key Laboratory of Quantum Information and Quantum Optoelectronic Devices,\\
School of Physics, Xi’an Jiaotong University, Xi’an 710049, China}

\author{Ren Zhang}
\affiliation{MOE Key Laboratory for Nonequilibrium Synthesis and Modulation of Condensed Matter,\\
Shaanxi Province Key Laboratory of Quantum Information and Quantum Optoelectronic Devices,\\
School of Physics, Xi’an Jiaotong University, Xi’an 710049, China}

\begin{abstract}
Non-Hermitian systems can evade Anderson localization and exhibit delocalized states even in one dimension. Here, we show that such non-Hermitian delocalized states under periodic boundary conditions (PBC) are intrinsically critical, realizing the universality class of one-dimensional random Dirac fermions. By linking spectral winding to topological Anderson transitions via Hermitization, we demonstrate that the delocalized PBC states exhibit a Dirac-type criticality with universal algebraic correlations. In contrast to Hermitian systems, where this criticality occurs only at fine-tuned transition points, it emerges generically in non-Hermitian systems as a consequence of spectral topology. These results identify a universal mechanism by which non-Hermiticity promotes criticality, providing a unified description of non-Hermitian delocalization in one dimension.
\end{abstract}

\maketitle

\textcolor{blue}{\textit{Introduction}---}Disordered Dirac Hamiltonians provide a fundamental route to quantum criticality in low-dimensional systems. In two dimensions, random Dirac fermions have been extensively studied, where disorder can lead to critical states with multifractal scaling and conformal invariance~\cite{Ludwig1991QHdirac, Chamon1996Multifractality, Carpentier2001critical, Evers2008RMPAndersonTransition,  Mudry1998RandomDirac}. In one dimension, random Dirac Hamiltonians also support critical states, but their critical properties are manifested differently. Rather than being characterized by conventional multifractal spectra, the criticality is primarily encoded in power-law correlations governed by an underlying Liouville-type field theory~\cite{Kogan1996randomDirac, Shelton1998LiouvilleQM}.  Such Dirac critical states typically appear only at finely tuned transition points, such as topological Anderson transitions~\cite{Li2009TopologicalAnderson, Groth2009TAI, Mondragon2014TopolocialCriticality}, rendering their realization in generic physical settings highly nontrivial.

In parallel, the discovery of the non-Hermitian skin effect (NHSE)—the extensive accumulation of eigenstates at system boundaries—has attracted significant attentions in recent years~\cite{Yao2018NHSE,yao2018chern,kunst2018biorthogonalBBC,Yokomizo2019NHSE,Xiao2020NHbulkboundary,Helbig2020generalBBC,Okuma2020SkinTopology,Zhang2020SpectralWinding,Lee2019NHSEanatomy,Ashida2021Non,Bergholtz2021RMP,Gohsrich2024Perspective,Wang2024Amoeba}. Beyond its unconventional bulk–boundary correspondence, the NHSE is rooted in a nontrivial spectral topology characterized by spectral winding in the complex-energy plane. In the presence of disorder, non-Hermitian systems exhibit a rich variety of phenomena~\cite{kawabata2026nonhermitiandisorderedsystems}, including topological Anderson phases~\cite{Zhang2020NHtopoAnderson,Tang2020TopoAnderson,Liu2021NHtopoAnderson2D,Lin2022experimentNHtopoAI},  many-body localization~\cite{Hamazaki2019nonHermitianMBL,Zhai2020MBLquasiperiodic,Suthar2022nhMBL,Wang2023skinMBL,Roccati2024nonHermitianMBLchaos}, unconventional spectral structures~\cite{Marchetti2001susyDOS,Silvestrov2001NHtailDOS,Longhi2025LifshitzTail,Hamazaki2020NHmatrixuniversality,Chen2025NHsigmaModel}, and anomalous dynamics~\cite{Balasubrahmaniyam2020Necklace,Weidemann2021NHtransport,Longhi2023LocLindblad,Yusipov2018AndersonJump, Tzortzakakis2021NHtransport, Leventis2022NHjump, Sahoo2022NHAndersonTransport,Tzortzakakis2020complexdisorder,Yusipov2017OpenSysLocalization,Li2025universaldynamics,Xing2025spreading,Shang2025spreading,Li2024nonblochdynamics}. A particularly striking consequence is the NHSE-driven Anderson transition, which enables delocalization even in one dimension—generically forbidden in Hermitian systems~\cite{Hatano1996HNmodel,Hatano1997HNmodel,Efetov1997DirectedChaos, Efetov1997DirectedDisorder,
Brouwer1997DirectedLocalization,
Hatano1998delocalization, Nelson1998NHbiology, Shnerb1998WindingNumber, Goldsheid1998NHAnderson, 
Feinberg1999NHspectrum, Feinberg1999delocalization, Kolesnikov2000delocTransition, Jiang2019NHSEquasiperiodic,Longhi2019TopoQuasicrystal,Longhi2019NHandersionTransition,Huang2020nhAnderson,Liu2020NHquasiperiodic,Zeng2020NHmobility,Kawabata2021NonunitaryScaling,Weidemann2022NHfloquetquasicrystal,Luo2021transfermatrixAT,Luo2021NHandersonTransition,Liu2021NHquasicrystal,Luo2022NHandersonTransition,Liu2024dissipationLocal,Wang2025nonBlochAT, Longhi2025erratic, Wang2025delocalization,Huang2025UniversalScaling, Wang2025delocalization3D,shang2025anisotropicLocalization,wang2026skinanderson,Longhi2021disorderNHSE,Sun2025skinAndersonTransition}. 


In this work, we show that non-Hermitian delocalization induced by the NHSE is in fact generically critical and belongs to the universality class of one-dimensional random Dirac fermions. In particular, the delocalized modes under periodic boundary conditions (PBC) realize critical states governed by an emergent random Dirac equation, with their statistical properties described by Liouville-type theory. More specifically, the correspondence between spectral winding and the topology of Hermitized Hamiltonian~\cite{Feinberg1997Hermitization} naturally links non-Hermitian delocalized states to critical states at a topological Anderson transition.  Based on this, we unveil the universal feature of the non-Hermitian delocalization: the delocalized eigenstates under PBC are critical and exhibit an algebraic spatial correlation,
\begin{eqnarray}\label{eq:critical_correlation}
\langle|\psi(x)\psi(0)|^q\rangle_{dis}\sim x^{-3/2}\qquad (x\gg 1),
\end{eqnarray}
which serves as a hallmark of the criticality of one-dimensional Dirac fermions. Here, $q$ labels the moment order and $\langle\cdots\rangle_{dis}$ denotes
disorder average. These results demonstrate that non-Hermitian systems provide a generic and robust route to realizing a distinct class of Dirac critical states that are otherwise accessible only at fine-tuned points in Hermitian settings.

\begin{figure}
    \centering
    \begin{tabular}{cc}
    \includegraphics[width=0.9\linewidth]{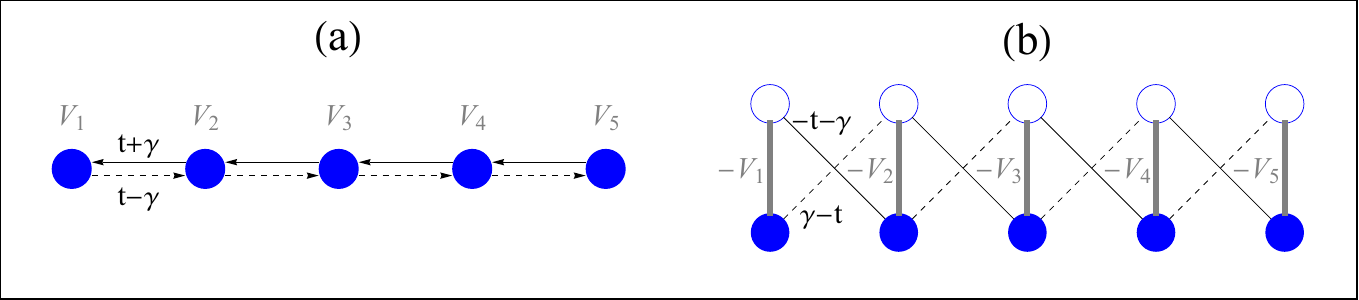}  \\
    \includegraphics[width=1\linewidth]{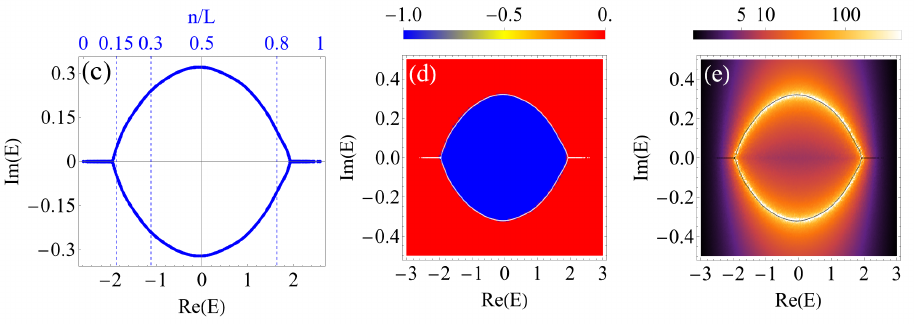} 
    \end{tabular}
    \caption{Schematic plot for the (a) Hatano-Nelson model and  (b) the Hermitization model ($E=0$ is assumed). (c) The PBC spectrum of the Hatano-Nelson model, where the upper ticks (blue) label the order of eigenstates according to the real part of their eigenvalues, where the order index is normalized by the system size (the total number of eigenstates). (d) The topological invariant of the Hamiltonian $\Tilde{H}$ as a function of $(E_r, E_i)$. (e) Localization length as a function of complex energy (extracted from a system with $L=2000$). In (d) and (e), both of the white and black lines show the overlaid PBC spectrum. In the plots, the parameters are $t=1$, $\gamma=0.2$, and $W=1$.}
    \label{fig:map}
\end{figure}

\textcolor{blue}{\textit{Spectral topology and topological Anderson transition}---}As a concrete example, we consider the Hatano–Nelson (HN) model [see Fig.~\ref{fig:map}(a)]:  
\begin{equation}\label{eq:HNmodel}
H_{HN}=\sum_{j=1}^L(t+\gamma)c^\dagger_j c_{j+1}+(t-\gamma)c^\dagger_{j+1}c_{j} +V_j c^\dagger_j c_j   
\end{equation}
where $c_{L+1}=c_1$ is assumed with $L$ being the system size, $t\pm\gamma$ describe asymmetric hopping, and $V_j$ is real-valued random variables following a box distribution $V_j\in[-W,W]$. It is known that the PBC spectrum of the HN model is a lemon shape attached with two wings, as shown in Fig.~\ref{fig:map} (b), where the loop part is constituted of delocalized modes (dub them loop states).

To unveil the connection between the loop states and the TAI critical states, we analyze the Hamiltonian $\Tilde{H}$ constructed via the Hermitization method~\cite{Feinberg1997Hermitization}, 
\begin{equation}\label{eq:Hermitzation}
\Tilde{H}=\left(\begin{array}{cc}
   0  & E-H_{HN} \\
  E^\ast-H_{HN}^\dagger   & 0
\end{array}\right),  
\end{equation}
which describes a TAI in the chiral class.  For the Hatano–Nelson model [Eq.~\eqref{eq:HNmodel}, Fig.~\ref{fig:map}(a)], $\tilde{H}$ defines a two-band model illustrated in Fig.~\ref{fig:map}(b); in the special case $t = \pm \gamma$, it reduces to a SSH model with random intra-cell couplings. A key observation is that if $|\psi\rangle$ and $|\tilde{\psi}\rangle$ respectively denote the right and left eigenstates of $H_{HN}$ corresponding the same energy $E$, i.e., $H_{HN}|\psi\rangle = E|\psi\rangle$ and $H^\dagger_{HN}|\tilde{\psi}\rangle =E^\ast|\Tilde{\psi}\rangle$, the two zero modes of $\tilde{H}$ are given by
\begin{eqnarray}\label{eq:zeromodes}
|\Psi_{0}^{(1)}\rangle \propto (0,|\psi\rangle)^T,\qquad|\Psi_{0}^{(2)}\rangle\propto(|\Tilde{\psi}\rangle,0)^T.
\end{eqnarray}
This establishes that a non-Hermitian skin mode of $H_{HN}$ corresponds to a boundary-localized topological edge mode of $\tilde{H}$. Consequently, the emergence of the skin mode and the topological edge mode are simultaneous: a nonzero PBC spectral winding of $H_{HN}$ is equivalent to a nonzero topological invariant (winding number) of $\tilde{H}$. Importantly, this correspondence holds even in the presence of disorder~\cite{Claes2021SkinDisorder, Supp}.

Treating $E = E_r+i E_i$ as a varying parameter for $\tilde{H}$, a phase boundary naturally emerges in the $(E_r, E_i)$ plane separating topologically trivial and nontrivial regions. For any $E$ within the topologically nontrivial region, the spectral winding number of $H_{HN}$ with respect to $E$ is necessarily nonzero. Consequently, the topologically nontrivial region on the complex-energy plane coincides with the area enclosed by the loop portion of the PBC spectrum of $H_{HN}$. To verify this, we plot the topological invariant (winding number in the chiral AIII class, see Ref.~\cite{Supp}) of $\Tilde{H}$ as a function of $(E_r, E_i)$ in Fig.~\ref{fig:map}(c). As expected, the topological phase boundary coincides precisely with the loop part of the PBC spectrum of $H_{HN}$.  



By connecting the phase boundary to the PBC spectrum, we demonstrate that all PBC loop states correspond precisely to the critical states of $\tilde{H}$ at the topological Anderson transition. These critical states feature a diverging localization length, which is mirrored by the divergence of the localization length of $H_{HN}$ along the PBC spectral loop [Fig.~\ref{fig:map}(e)], further confirming the correspondence.

\textcolor{blue}{\textit{Delocalization and disordered Dirac Hamiltonian}---}Using this correspondence, the spatial structure of the loop-state wavefunctions can be inferred from the critical states at the topological Anderson transition of $\tilde{H}$. In the absence of disorder, $\tilde{H}$ in momentum space takes the form $\Tilde{H}=(E_r-2t\cos k)\sigma^x+(E_i-2\gamma\sin k)\sigma^y$,
where $\sigma^x,\sigma^y$ act on sublattice degree of freedoms. The Hamiltonian is gapless at $k_0$ satisfying $E_r = 2t \cos k_0$ and $E_i = 2\gamma \sin k_0$. Including disorder, a long-wavelength expansion around $k_0$ yields 
\begin{equation}\label{eq:RandomDirac}
\Tilde{H}_{eff}=(\eta_x\sigma^x+\eta_y\sigma^y)(-i\partial_x)+m(x)\sigma^x   
\end{equation}
where $\eta_x=t E_i/\gamma$, $\eta_y=\gamma E_r/t$, and $m(x)$ represents the long-wavelength component of the on-site potential $V_j$, satisfying $\langle m(x) m(x^\prime)\rangle_{dis}=g\delta(x-x^\prime)$, where $g$ measures the disorder strength. Equation~\eqref{eq:RandomDirac} thus describes a Dirac Hamiltonian with a random mass term. The zero-energy critical state near the phase boundary then takes the ansatz
\begin{equation}\label{eq:zeromode}
\Psi_0(x)=\frac{1}{\sqrt{\mathcal N}}\Vec{\chi} \exp [\alpha V(x)],\quad    V(x)=\int_0^x dym(y),    
\end{equation}
where $\Vec{\chi}$ is a spinor in the sublattice space, $\alpha=\alpha_r+i\alpha_i$, and $\mathcal N=\int_{-L/2}^{L/2} dx\exp [2\alpha_rV(x)]$ is the normalization factor. To ensure the ansatz remains valid, the coefficient of $m(x)$ in the eigenvalue equation must vanish, yielding 
\begin{equation}
[(1-i\alpha\eta_x)\sigma^x-i\alpha\eta_y\sigma^y]\Vec{\chi}=0, 
\end{equation}
which is solved by $\alpha_1=(\eta_y-i\eta_x)/(\eta_x^2+\eta^2_y)$ or $\alpha_2=(-\eta_y-i\eta_x)/(\eta_x^2+\eta^2_y)=-\alpha_1^\ast$. The corresponding eigenvectors are  $\vec{\chi}_1=(0,1)^T$ and $\vec{\chi}_2=(1,0)^T$, consistent with Eq.~\eqref{eq:zeromodes}.
Therefore, in the long-wavelength limit, the right eigenstate of $H_{HN}$ with energy $E$ is given by
\begin{eqnarray}\label{eq:longwavefunc}
\psi(x)=\frac{1}{\sqrt{\mathcal N}}\exp [\alpha_1 V(x)],  
\end{eqnarray}
where the prefactor is settled to analyze the localization properties, for example, the participation ratio. Below, we will focus on the right eigenstate,  and all the relevant results hold for the left eigenstate accordingly. 

After establishing the long-wavelength form Eq.~\eqref{eq:longwavefunc} of the loop states from the correspondence, we expect the wavefunction to inherit the critical feature of the TAT states. Below, we examine the localization properties of the loop states by combining numerics from the lattice model Eq.~\eqref{eq:HNmodel} and analytics from Eq.~\eqref{eq:longwavefunc}. 




\begin{figure}
    \centering
    \includegraphics[width=1\linewidth]{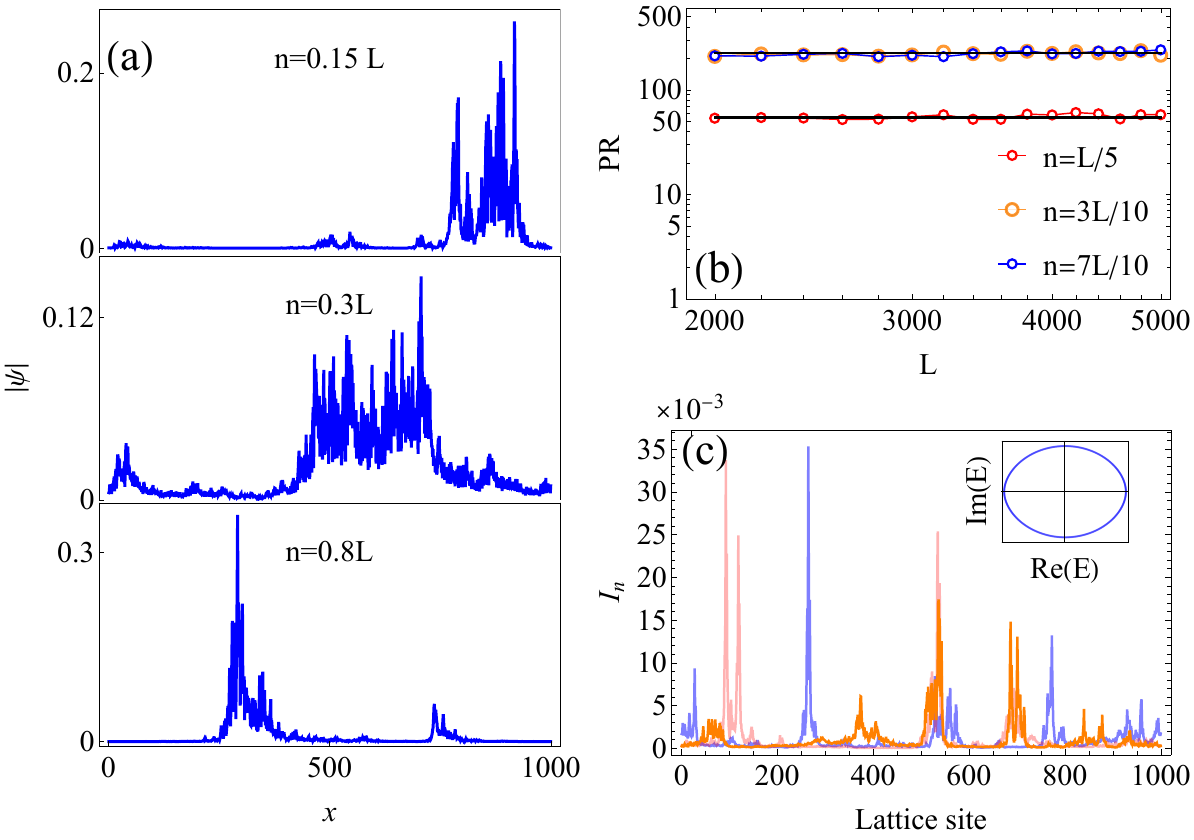}  
    \caption{(a) The wavefunction profile of selected eigenstates by the order of their real part of eigenvalues [see Fig.~\ref{fig:map} (c)] under PBC.   (c) Participation ratio for selected eigenstates as a function of system size. Here, an average is taken over 200 realizations of disorder. (a) and (b) take the same parameters as Fig.~\ref{fig:map}. (d) Wave intensity profile for different disorder realizations, where parameters $t=1$, $\gamma=0.8$, and $W=1$ are used; the inset shows the corresponding PBC spectrum.}
    \label{fig:spectr_wave}
\end{figure}

\textcolor{blue}{\textit{Spatial distribution of delocalized modes}---}In Fig.~\ref{fig:spectr_wave}(a), selected PBC loop states are plotted; their profiles are largely confined to a finite region, distinct from typical extended or exponentially-localized waves~\cite{Silvestrov1998localization, Hatano1998eigenfunctions}. To quantify their localization properties, we evaluate the participation ratio (PR) of the right eigenstate $|\psi\rangle$:
\begin{eqnarray}
\text{PR}=\Big(\sum_{j=1}^L|\psi_j|^2\Big)^2\Big/\sum_{j=1}^L |\psi_j|^4.    
\end{eqnarray}
For an extended state, the PR is expected to grow with system size, whereas it saturates for a localized state. As shown in Fig.~\ref{fig:spectr_wave}(b), the PRs of the selected loop states remain essentially size-independent for sufficiently large systems, strongly implying a localized nature. This seems inconsistent with prior findings that their localization length diverges~\cite{Claes2021SkinDisorder} [see also Fig.~\ref{fig:map}(e)]. This contradiction can be resolved by realizing the critical nature of these states as discussed later.

To further characterize their spatial structure, we analyze the average wavefunction distribution for all states on the complex-energy loop:
\begin{eqnarray}
I_j=\frac{1}{N_{loop}}\sum_{\varepsilon_\alpha\in loop}|\psi_j(\varepsilon_\alpha)|^2,    
\end{eqnarray}
where $j$ indexes the lattice site, $\alpha$ labels different eigenstates, and $N_{loop}$ counts the sates lying on the spectral loop~\cite{Longhi2025ErraticSkin}. For a given disorder realization, one expects $I_j$ to be nearly uniform for extended states or for exponentially localized states with centers uniformly distributed in space. However, as shown in Fig.~\ref{fig:spectr_wave}(c), $I_j$ for three disorder realizations exhibits a randomly localized pattern, resembling the so-called erratic skin modes~\cite{Longhi2025ErraticSkin}. This clearly rules out the possibility that the loop states are extended or exponentially localized. In these plots, we use a strong nonreciprocal parameter $\gamma$ such that all eigenstates are promoted to loop states.

The disorder-configuration sensitive behavior of $I_j$ can be understood from the profile $ e^{\alpha_1 V(x)}$ in Eq.~\eqref{eq:longwavefunc}, which is fully determined by the random mass term $m(x)$. The mass $m(x)$ effectively matches the long-wavelength structure of the local Hamiltonian along the spectral loop and varies slowly for neighboring states. Consequently, eigenstates located closely on the loop exhibit similar ~\cite{Supp}. Depending on the disorder configuration, this leads the loop states to cluster around a few spatial regions, producing a highly non-uniform distribution.

\begin{figure}
    \centering \includegraphics[width=1\linewidth]{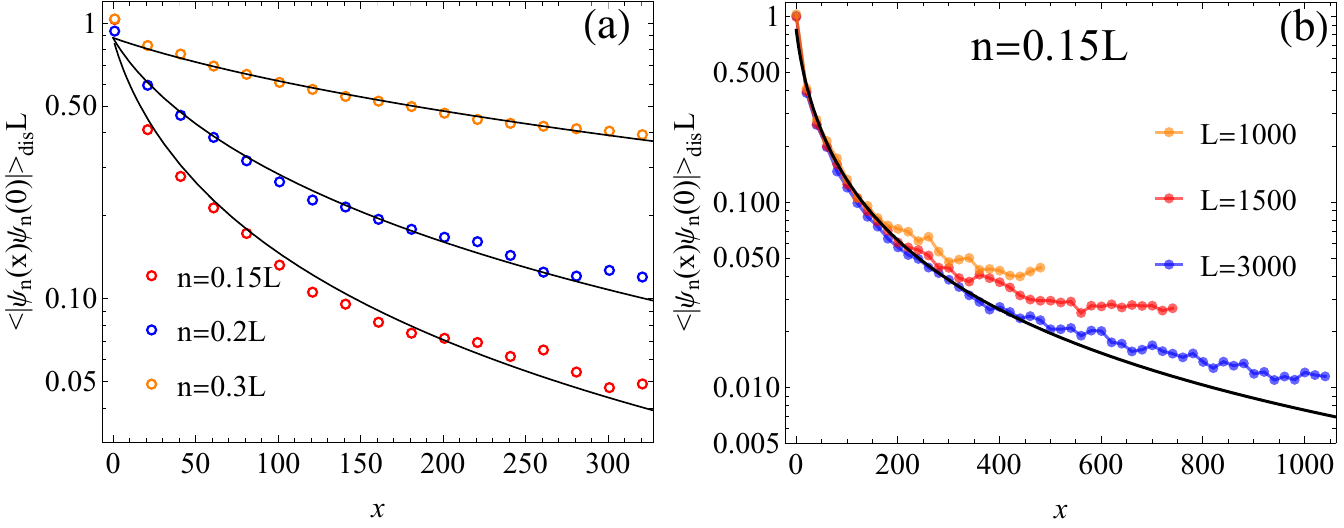} 
    \caption{(a) Wavefunction correlation as a function of distance, where $L=1000$ and $5\times 10^3$ realizations of disorder are performed. (b) the correlation for $``n=0.15 L"$-th state for three systems with different sizes, where $5\times 10^3-2\times 10^5$ disorder configurations are simulated.
    Other parameters are $t=1$, $\gamma=0.2$, and $W=1$. In both plots, the solid lines represent the fitting function $I(x,q)$ where $x$ is appropriately rescaled.}
    \label{fig:correlation}
\end{figure}

\textcolor{blue}{\textit{Criticality from correlation}---}The critical properties of $\psi(x)$ can be characterized through the correlations of the wavefunction profiles~\cite{Kogan1996randomDirac, Shelton1998LiouvilleQM,Blents1997delocalization}:
\begin{eqnarray}\label{eq:zeromode_correlation}
W_q(x,L)&&=\langle|\psi(x)\psi(0)|^q\rangle_{dis}\nonumber\\
&&=\Big\langle\Big|\frac{1}{\sqrt{\mathcal N}}e^{V(x)}\frac{1}{\sqrt{\mathcal N}}e^{V(0)}\Big|^q\Big\rangle_{dis},
\end{eqnarray}
where we set $\alpha_1=1$ because only Re$(\alpha_1)$ is relevant and can be absorbed into $V(x)$. Using the Kogan–Mudry–Tsvelik approach~\cite{Kogan1996randomDirac, Shelton1998LiouvilleQM, Supp}, the correlation can be converted to a statistics problem dictated by a Liouville Hamiltonian
\begin{eqnarray}
H=-\frac{g}{2}\frac{d^2}{dV^2}+\omega e^{2V},    
\end{eqnarray}
which represents a particle moving in an exponential potential, with $V$ serving as the spatial variable. Following Ref.~\cite{Blents1997delocalization} and applying a hard-wall approximation, the correlation can be calculated analytically
\begin{eqnarray}\label{eq:correlationresult1}
W_q(x,L)=\frac{32 q^2}{\Gamma(q)\pi L}I(x,q),
\end{eqnarray}
where $I(x,q)=\int_0^\infty dk\frac{k^2e^{-xk^2}}{(q^2+k^2)^4}$. For a large distance $x\gg 1$, $I(x,q)=\sqrt{\pi}/(4q^8 x^{3/2})+O(x^{-5/2})\sim x^{-3/2}$, so that the wavefunction correlation behaves as $ W_q(x,L)\sim x^{-3/2}$, i.e., the main result in  Eq.~\eqref{eq:critical_correlation},
which is independent of $q$. 
Here, the power-law decay $x^{-3/2}$ is a manifestation of criticality, sharply differing from the exponential decay of conventional Anderson-localized states.

\begin{figure}
    \centering
    \begin{tabular}{cc} \includegraphics[width=1\linewidth]{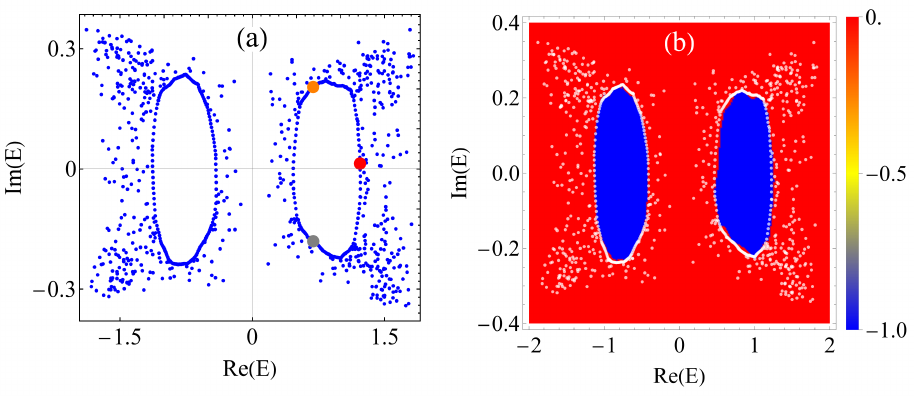}\\     \includegraphics[width=0.95\linewidth]{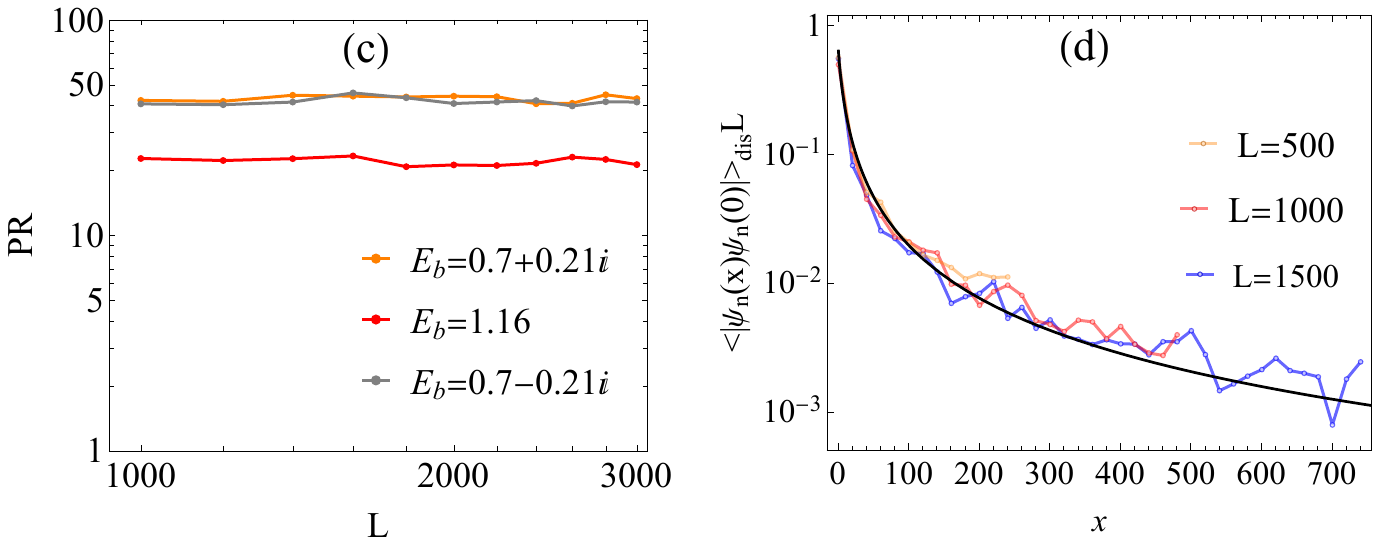}
    \end{tabular}
    \caption{(a) The spectrum of the two-band model with disorder. (b) The topological invariant of the corresponding $\tilde{H}$, with the white dots indicating the overlaid PBC spectrum. (c) The averaged participation ratio for eigenstates on the spectral loop with energies corresponding to the three makers in (a) (in the same colors). (d) The wave function correlations for the loop state with energy near the red marker in (a). In the plots, parameters are: $r=1, v=0.3, W=1$, and $\lambda=0.8$. }
    \label{fig:two_band}
\end{figure}


To validate our theoretical prediction, in Fig.~\ref{fig:correlation}, we numerically calculated the correlations of loop states in the Hatano-Nelson model [Eq.~\eqref{eq:HNmodel}] under PBC. Here, only $x \leq L/2$ is considered under PBC.
The numerical data agree well with Eq.~\eqref{eq:correlationresult1} after rescaling the distance $x$, reflecting that Eq.~\eqref{eq:correlationresult1} is derived in the long-wavelength limit, where the lattice unit length is not explicitly defined.
Verifying the scaling in Eq.~\eqref{eq:critical_correlation} requires exceptionally large system sizes, which are challenging to access numerically. Nevertheless, Fig.~\ref{fig:correlation}(b) shows that as the system size increases, the data at longer correlation distances increasingly approach the prediction of Eq.~\eqref{eq:correlationresult1}. In the large-$L$ limit, the $x^{-3/2}$ scaling is expected to emerge fully.

The correlation Eq.~\eqref{eq:correlationresult1} can also be used to understand the size-independent participation ratio (PR) in Fig.~\ref{fig:spectr_wave} (b). For $x=0$, $I(0,q)=\pi/(32 q^5)$, yielding $\langle|\psi_j|^{2q}\rangle_{dis}\sim W(0,L)\sim L^{-1}$, given that $|\psi\rangle$ is normalized. Here, the position-independent value is because translational symmetry is restored after disorder average. The PR then scales as $\text{PR}\sim [L\langle|\psi_j|^{4}\rangle_{dis}]^{-1}\sim 1$, explaining its apparent independence from system size, in contrast to the multifractal scaling of random Dirac fermions in two dimensions~\cite{Evers2008RMPAndersonTransition}.






\textcolor{blue}{\textit{The universality}---}While we have focused on the Hatano–Nelson model, the critical nature of the loop states under PBC is universal for disordered systems exhibiting NHSE. This universality arises from the generic mapping between loop states and the critical states at TAT. Below, we discuss other scenarios that fall within this framework.

\textbf{Complex-valued disorder.} A natural extension of the Hatano–Nelson model replaces the on-site potential with complex-valued disorder, i.e., $V_j\rightarrow u_j+iv_j$ where $u_j,v_j$ are independent random variables. Numerical tests confirm that the loop eigenstates retain the same properties~\cite{Supp}. In the Hermitization framework, the long-wavelength expansion yields a random Dirac Hamiltonian,
$\Tilde{H}_{eff}=(\eta_x\sigma^x+\eta_y\sigma^y)(-i\partial_x)+u(x)\sigma^x +v(x)\sigma^y$, where $u(x)$ and $v(x)$ are Gaussian random fields. A possible zero mode has the profile $|\Psi_0(x)|\sim e^{\int_0^x dy[\alpha_{r}u(y)-\alpha_{i} v(y)]}$ with solvable coefficients $\alpha_r$ and $\alpha_i$. By analogy with Eq.~\eqref{eq:zeromode}, the correlation functions in Eqs.~\eqref{eq:correlationresult1} and~\eqref{eq:critical_correlation} are expected to remain valid.

\textbf{A tow-band system.} The NHSE can also arise from local gain and loss, such as the following  two-band Hamiltonian~\cite{Lee2016edgeModes,Yao2018NHSE} 
\begin{eqnarray}
H_{two-band}= (v+r\cos k)\tau^x+(r\sin k+i\frac{\lambda}{2})\tau^z.    
\end{eqnarray}
On-site disorder is added as $H_{dis}=\sum_{j=1}^{2L} V_jc_j^\dagger c_j$,
where index ``$j$" runs over $2L$ sites in $L$ unit cells, and $V_j$ are independent random variables uniformly distributed in $[-W, W]$, regardless of sublattice. Figure~\ref{fig:two_band} shows the spectrum and the topological phase diagram of the Hermitized Hamiltonian $\tilde{H}$ in the complex-energy plane, where delocalized states reside on the phase boundary, along with the PR and wavefunction correlations. Notably, the wavefunction correlation shows only a weak system-size dependence, likely due to the stronger localization of the states; nevertheless, it still obeys the same scaling predicted by Eq.~\eqref{eq:correlationresult1}. This case shares the same effective long-wavelength theory as the Hatano-Nelson model with complex disorder discussed above~\cite{Supp}.

\textbf{Random hopping models.} Another intriguing scenario arises when the skin modes are entirely induced by disorder, i.e., the non-reciprocity originates purely from bond disorder~\cite{Claes2021SkinDisorder, Longhi2020RandomSkin}. In this case, the mapping between PBC loop states and topological transition modes still holds, and analogous wavefunction properties are expected~\cite{Claes2021SkinDisorder}.

A special case worth mentioning is the so-called erratic skin effect~\cite{Longhi2025ErraticSkin}, where the asymmetric forward ($e^{V_j}$) and backward ($e^{-V_j}$) hopping are associated with the same random variable $V_j$. The resultant eigenmodes follow a profile of the form in Eq.~\eqref{eq:zeromode}, with the exponent accumulating along a random-walk trajectory. Consequently, the states obey the same correlations as in Eqs.~\eqref{eq:correlationresult1} and~\eqref{eq:critical_correlation}. However, these modes arise from a special construction rather than from non-Hermitian topology. 


\textcolor{blue}{\textit{Conclusions}---}We have shown that non-Hermitian delocalization in one dimension is generically critical rather than extended, and belongs to the universality class of random Dirac fermions. While such criticality in Hermitian systems appears only at finely tuned transition points, here it emerges generically along spectral loops due to non-Hermitian spectral topology. This correspondence between spectral winding and topological Anderson transitions identifies the loop states as Dirac critical states with universal algebraic correlations, providing a unified physical picture of non-Hermitian delocalization. Our results establish non-Hermitian systems as a natural platform for realizing the criticality of Dirac fermions in one dimension without fine tuning.

\textcolor{blue}{\textit{Acknowledgements}---} We thank Prof. Zhong Wang for insightful discussions. B.L. acknowledges support from the NSFC under Grant No.~12404185. S.Z. and R.Z. are supported by the NSFC under Grant No.~12574299.

\bibliographystyle{apsrev4-1-title}
\bibliography{delocalization}

\begin{thebibliography}{94}%
\makeatletter
\providecommand \@ifxundefined [1]{%
 \@ifx{#1\undefined}
}%
\providecommand \@ifnum [1]{%
 \ifnum #1\expandafter \@firstoftwo
 \else \expandafter \@secondoftwo
 \fi
}%
\providecommand \@ifx [1]{%
 \ifx #1\expandafter \@firstoftwo
 \else \expandafter \@secondoftwo
 \fi
}%
\providecommand \natexlab [1]{#1}%
\providecommand \enquote  [1]{``#1''}%
\providecommand \bibnamefont  [1]{#1}%
\providecommand \bibfnamefont [1]{#1}%
\providecommand \citenamefont [1]{#1}%
\providecommand \href@noop [0]{\@secondoftwo}%
\providecommand \href [0]{\begingroup \@sanitize@url \@href}%
\providecommand \@href[1]{\@@startlink{#1}\@@href}%
\providecommand \@@href[1]{\endgroup#1\@@endlink}%
\providecommand \@sanitize@url [0]{\catcode `\\12\catcode `\$12\catcode `\&12\catcode `\#12\catcode `\^12\catcode `\_12\catcode `\%12\relax}%
\providecommand \@@startlink[1]{}%
\providecommand \@@endlink[0]{}%
\providecommand \url  [0]{\begingroup\@sanitize@url \@url }%
\providecommand \@url [1]{\endgroup\@href {#1}{\urlprefix }}%
\providecommand \urlprefix  [0]{URL }%
\providecommand \Eprint [0]{\href }%
\providecommand \doibase [0]{http://dx.doi.org/}%
\providecommand \selectlanguage [0]{\@gobble}%
\providecommand \bibinfo  [0]{\@secondoftwo}%
\providecommand \bibfield  [0]{\@secondoftwo}%
\providecommand \translation [1]{[#1]}%
\providecommand \BibitemOpen [0]{}%
\providecommand \bibitemStop [0]{}%
\providecommand \bibitemNoStop [0]{.\EOS\space}%
\providecommand \EOS [0]{\spacefactor3000\relax}%
\providecommand \BibitemShut  [1]{\csname bibitem#1\endcsname}%
\let\auto@bib@innerbib\@empty
\bibitem [{\citenamefont {Ludwig}\ \emph {et~al.}(1994)\citenamefont {Ludwig}, \citenamefont {Fisher}, \citenamefont {Shankar},\ and\ \citenamefont {Grinstein}}]{Ludwig1991QHdirac}%
  \BibitemOpen
  \bibfield  {author} {\bibinfo {author} {\bibfnamefont {A.~W.~W.}\ \bibnamefont {Ludwig}}, \bibinfo {author} {\bibfnamefont {M.~P.~A.}\ \bibnamefont {Fisher}}, \bibinfo {author} {\bibfnamefont {R.}~\bibnamefont {Shankar}}, \ and\ \bibinfo {author} {\bibfnamefont {G.}~\bibnamefont {Grinstein}},\ }\bibfield  {title} {\enquote {\bibinfo {title} {Integer quantum hall transition: An alternative approach and exact results},}\ }\href {\doibase 10.1103/PhysRevB.50.7526} {\bibfield  {journal} {\bibinfo  {journal} {Phys. Rev. B}\ }\textbf {\bibinfo {volume} {50}},\ \bibinfo {pages} {7526} (\bibinfo {year} {1994})}\BibitemShut {NoStop}%
\bibitem [{\citenamefont {Chamon}\ \emph {et~al.}(1996)\citenamefont {Chamon}, \citenamefont {Mudry},\ and\ \citenamefont {Wen}}]{Chamon1996Multifractality}%
  \BibitemOpen
  \bibfield  {author} {\bibinfo {author} {\bibfnamefont {C.~d.~C.}\ \bibnamefont {Chamon}}, \bibinfo {author} {\bibfnamefont {C.}~\bibnamefont {Mudry}}, \ and\ \bibinfo {author} {\bibfnamefont {X.-G.}\ \bibnamefont {Wen}},\ }\bibfield  {title} {\enquote {\bibinfo {title} {Localization in two dimensions, gaussian field theories, and multifractality},}\ }\href {\doibase 10.1103/PhysRevLett.77.4194} {\bibfield  {journal} {\bibinfo  {journal} {Phys. Rev. Lett.}\ }\textbf {\bibinfo {volume} {77}},\ \bibinfo {pages} {4194} (\bibinfo {year} {1996})}\BibitemShut {NoStop}%
\bibitem [{\citenamefont {Carpentier}\ and\ \citenamefont {Le~Doussal}(2001)}]{Carpentier2001critical}%
  \BibitemOpen
  \bibfield  {author} {\bibinfo {author} {\bibfnamefont {D.}~\bibnamefont {Carpentier}}\ and\ \bibinfo {author} {\bibfnamefont {P.}~\bibnamefont {Le~Doussal}},\ }\bibfield  {title} {\enquote {\bibinfo {title} {Glass transition of a particle in a random potential, front selection in nonlinear renormalization group, and entropic phenomena in liouville and sinh-gordon models},}\ }\href {\doibase 10.1103/PhysRevE.63.026110} {\bibfield  {journal} {\bibinfo  {journal} {Phys. Rev. E}\ }\textbf {\bibinfo {volume} {63}},\ \bibinfo {pages} {026110} (\bibinfo {year} {2001})}\BibitemShut {NoStop}%
\bibitem [{\citenamefont {Evers}\ and\ \citenamefont {Mirlin}(2008)}]{Evers2008RMPAndersonTransition}%
  \BibitemOpen
  \bibfield  {author} {\bibinfo {author} {\bibfnamefont {F.}~\bibnamefont {Evers}}\ and\ \bibinfo {author} {\bibfnamefont {A.~D.}\ \bibnamefont {Mirlin}},\ }\bibfield  {title} {\enquote {\bibinfo {title} {Anderson transitions},}\ }\href {\doibase 10.1103/RevModPhys.80.1355} {\bibfield  {journal} {\bibinfo  {journal} {Rev. Mod. Phys.}\ }\textbf {\bibinfo {volume} {80}},\ \bibinfo {pages} {1355} (\bibinfo {year} {2008})}\BibitemShut {NoStop}%
\bibitem [{\citenamefont {Mudry}\ \emph {et~al.}(1998)\citenamefont {Mudry}, \citenamefont {Simons},\ and\ \citenamefont {Altland}}]{Mudry1998RandomDirac}%
  \BibitemOpen
  \bibfield  {author} {\bibinfo {author} {\bibfnamefont {C.}~\bibnamefont {Mudry}}, \bibinfo {author} {\bibfnamefont {B.~D.}\ \bibnamefont {Simons}}, \ and\ \bibinfo {author} {\bibfnamefont {A.}~\bibnamefont {Altland}},\ }\bibfield  {title} {\enquote {\bibinfo {title} {Random dirac fermions and non-hermitian quantum mechanics},}\ }\href {\doibase 10.1103/PhysRevLett.80.4257} {\bibfield  {journal} {\bibinfo  {journal} {Phys. Rev. Lett.}\ }\textbf {\bibinfo {volume} {80}},\ \bibinfo {pages} {4257} (\bibinfo {year} {1998})}\BibitemShut {NoStop}%
\bibitem [{\citenamefont {Kogan}\ \emph {et~al.}(1996)\citenamefont {Kogan}, \citenamefont {Mudry},\ and\ \citenamefont {Tsvelik}}]{Kogan1996randomDirac}%
  \BibitemOpen
  \bibfield  {author} {\bibinfo {author} {\bibfnamefont {I.~I.}\ \bibnamefont {Kogan}}, \bibinfo {author} {\bibfnamefont {C.}~\bibnamefont {Mudry}}, \ and\ \bibinfo {author} {\bibfnamefont {A.~M.}\ \bibnamefont {Tsvelik}},\ }\bibfield  {title} {\enquote {\bibinfo {title} {Liouville theory as a model for prelocalized states in disordered conductors},}\ }\href {\doibase 10.1103/PhysRevLett.77.707} {\bibfield  {journal} {\bibinfo  {journal} {Phys. Rev. Lett.}\ }\textbf {\bibinfo {volume} {77}},\ \bibinfo {pages} {707} (\bibinfo {year} {1996})}\BibitemShut {NoStop}%
\bibitem [{\citenamefont {Shelton}\ and\ \citenamefont {Tsvelik}(1998)}]{Shelton1998LiouvilleQM}%
  \BibitemOpen
  \bibfield  {author} {\bibinfo {author} {\bibfnamefont {D.~G.}\ \bibnamefont {Shelton}}\ and\ \bibinfo {author} {\bibfnamefont {A.~M.}\ \bibnamefont {Tsvelik}},\ }\bibfield  {title} {\enquote {\bibinfo {title} {Effective theory for midgap states in doped spin-ladder and spin-peierls systems: Liouville quantum mechanics},}\ }\href {\doibase 10.1103/PhysRevB.57.14242} {\bibfield  {journal} {\bibinfo  {journal} {Phys. Rev. B}\ }\textbf {\bibinfo {volume} {57}},\ \bibinfo {pages} {14242} (\bibinfo {year} {1998})}\BibitemShut {NoStop}%
\bibitem [{\citenamefont {Li}\ \emph {et~al.}(2009)\citenamefont {Li}, \citenamefont {Chu}, \citenamefont {Jain},\ and\ \citenamefont {Shen}}]{Li2009TopologicalAnderson}%
  \BibitemOpen
  \bibfield  {author} {\bibinfo {author} {\bibfnamefont {J.}~\bibnamefont {Li}}, \bibinfo {author} {\bibfnamefont {R.-L.}\ \bibnamefont {Chu}}, \bibinfo {author} {\bibfnamefont {J.~K.}\ \bibnamefont {Jain}}, \ and\ \bibinfo {author} {\bibfnamefont {S.-Q.}\ \bibnamefont {Shen}},\ }\bibfield  {title} {\enquote {\bibinfo {title} {Topological anderson insulator},}\ }\href {\doibase 10.1103/PhysRevLett.102.136806} {\bibfield  {journal} {\bibinfo  {journal} {Phys. Rev. Lett.}\ }\textbf {\bibinfo {volume} {102}},\ \bibinfo {pages} {136806} (\bibinfo {year} {2009})}\BibitemShut {NoStop}%
\bibitem [{\citenamefont {Groth}\ \emph {et~al.}(2009)\citenamefont {Groth}, \citenamefont {Wimmer}, \citenamefont {Akhmerov}, \citenamefont {Tworzyd\l{}o},\ and\ \citenamefont {Beenakker}}]{Groth2009TAI}%
  \BibitemOpen
  \bibfield  {author} {\bibinfo {author} {\bibfnamefont {C.~W.}\ \bibnamefont {Groth}}, \bibinfo {author} {\bibfnamefont {M.}~\bibnamefont {Wimmer}}, \bibinfo {author} {\bibfnamefont {A.~R.}\ \bibnamefont {Akhmerov}}, \bibinfo {author} {\bibfnamefont {J.}~\bibnamefont {Tworzyd\l{}o}}, \ and\ \bibinfo {author} {\bibfnamefont {C.~W.~J.}\ \bibnamefont {Beenakker}},\ }\bibfield  {title} {\enquote {\bibinfo {title} {Theory of the topological anderson insulator},}\ }\href {\doibase 10.1103/PhysRevLett.103.196805} {\bibfield  {journal} {\bibinfo  {journal} {Phys. Rev. Lett.}\ }\textbf {\bibinfo {volume} {103}},\ \bibinfo {pages} {196805} (\bibinfo {year} {2009})}\BibitemShut {NoStop}%
\bibitem [{\citenamefont {Mondragon-Shem}\ \emph {et~al.}(2014)\citenamefont {Mondragon-Shem}, \citenamefont {Hughes}, \citenamefont {Song},\ and\ \citenamefont {Prodan}}]{Mondragon2014TopolocialCriticality}%
  \BibitemOpen
  \bibfield  {author} {\bibinfo {author} {\bibfnamefont {I.}~\bibnamefont {Mondragon-Shem}}, \bibinfo {author} {\bibfnamefont {T.~L.}\ \bibnamefont {Hughes}}, \bibinfo {author} {\bibfnamefont {J.}~\bibnamefont {Song}}, \ and\ \bibinfo {author} {\bibfnamefont {E.}~\bibnamefont {Prodan}},\ }\bibfield  {title} {\enquote {\bibinfo {title} {Topological criticality in the chiral-symmetric aiii class at strong disorder},}\ }\href {\doibase 10.1103/PhysRevLett.113.046802} {\bibfield  {journal} {\bibinfo  {journal} {Phys. Rev. Lett.}\ }\textbf {\bibinfo {volume} {113}},\ \bibinfo {pages} {046802} (\bibinfo {year} {2014})}\BibitemShut {NoStop}%
\bibitem [{\citenamefont {Yao}\ and\ \citenamefont {Wang}(2018)}]{Yao2018NHSE}%
  \BibitemOpen
  \bibfield  {author} {\bibinfo {author} {\bibfnamefont {S.}~\bibnamefont {Yao}}\ and\ \bibinfo {author} {\bibfnamefont {Z.}~\bibnamefont {Wang}},\ }\bibfield  {title} {\enquote {\bibinfo {title} {Edge states and topological invariants of non-hermitian systems},}\ }\href {\doibase 10.1103/PhysRevLett.121.086803} {\bibfield  {journal} {\bibinfo  {journal} {Phys. Rev. Lett.}\ }\textbf {\bibinfo {volume} {121}},\ \bibinfo {pages} {086803} (\bibinfo {year} {2018})}\BibitemShut {NoStop}%
\bibitem [{\citenamefont {Yao}\ \emph {et~al.}(2018)\citenamefont {Yao}, \citenamefont {Song},\ and\ \citenamefont {Wang}}]{yao2018chern}%
  \BibitemOpen
  \bibfield  {author} {\bibinfo {author} {\bibfnamefont {S.}~\bibnamefont {Yao}}, \bibinfo {author} {\bibfnamefont {F.}~\bibnamefont {Song}}, \ and\ \bibinfo {author} {\bibfnamefont {Z.}~\bibnamefont {Wang}},\ }\bibfield  {title} {\enquote {\bibinfo {title} {Non-{{Hermitian Chern Bands}}},}\ }\href {\doibase 10.1103/PhysRevLett.121.136802} {\bibfield  {journal} {\bibinfo  {journal} {Phys. Rev. Lett.}\ }\textbf {\bibinfo {volume} {121}},\ \bibinfo {pages} {136802} (\bibinfo {year} {2018})}\BibitemShut {NoStop}%
\bibitem [{\citenamefont {Kunst}\ \emph {et~al.}(2018)\citenamefont {Kunst}, \citenamefont {Edvardsson}, \citenamefont {Budich},\ and\ \citenamefont {Bergholtz}}]{kunst2018biorthogonalBBC}%
  \BibitemOpen
  \bibfield  {author} {\bibinfo {author} {\bibfnamefont {F.~K.}\ \bibnamefont {Kunst}}, \bibinfo {author} {\bibfnamefont {E.}~\bibnamefont {Edvardsson}}, \bibinfo {author} {\bibfnamefont {J.~C.}\ \bibnamefont {Budich}}, \ and\ \bibinfo {author} {\bibfnamefont {E.~J.}\ \bibnamefont {Bergholtz}},\ }\bibfield  {title} {\enquote {\bibinfo {title} {Biorthogonal bulk-boundary correspondence in non-hermitian systems},}\ }\href {\doibase 10.1103/PhysRevLett.121.026808} {\bibfield  {journal} {\bibinfo  {journal} {Phys. Rev. Lett.}\ }\textbf {\bibinfo {volume} {121}},\ \bibinfo {pages} {026808} (\bibinfo {year} {2018})}\BibitemShut {NoStop}%
\bibitem [{\citenamefont {Yokomizo}\ and\ \citenamefont {Murakami}(2019)}]{Yokomizo2019NHSE}%
  \BibitemOpen
  \bibfield  {author} {\bibinfo {author} {\bibfnamefont {K.}~\bibnamefont {Yokomizo}}\ and\ \bibinfo {author} {\bibfnamefont {S.}~\bibnamefont {Murakami}},\ }\bibfield  {title} {\enquote {\bibinfo {title} {Non-bloch band theory of non-hermitian systems},}\ }\href {\doibase 10.1103/PhysRevLett.123.066404} {\bibfield  {journal} {\bibinfo  {journal} {Phys. Rev. Lett.}\ }\textbf {\bibinfo {volume} {123}},\ \bibinfo {pages} {066404} (\bibinfo {year} {2019})}\BibitemShut {NoStop}%
\bibitem [{\citenamefont {Xiao}\ \emph {et~al.}(2020)\citenamefont {Xiao}, \citenamefont {Deng}, \citenamefont {Wang}, \citenamefont {Zhu}, \citenamefont {Wang}, \citenamefont {Yi},\ and\ \citenamefont {Xue}}]{Xiao2020NHbulkboundary}%
  \BibitemOpen
  \bibfield  {author} {\bibinfo {author} {\bibfnamefont {L.}~\bibnamefont {Xiao}}, \bibinfo {author} {\bibfnamefont {T.}~\bibnamefont {Deng}}, \bibinfo {author} {\bibfnamefont {K.}~\bibnamefont {Wang}}, \bibinfo {author} {\bibfnamefont {G.}~\bibnamefont {Zhu}}, \bibinfo {author} {\bibfnamefont {Z.}~\bibnamefont {Wang}}, \bibinfo {author} {\bibfnamefont {W.}~\bibnamefont {Yi}}, \ and\ \bibinfo {author} {\bibfnamefont {P.}~\bibnamefont {Xue}},\ }\bibfield  {title} {\enquote {\bibinfo {title} {Non-hermitian bulk--boundary correspondence in quantum dynamics},}\ }\href {\doibase 10.1038/s41567-020-0836-6} {\bibfield  {journal} {\bibinfo  {journal} {Nature Physics}\ }\textbf {\bibinfo {volume} {16}},\ \bibinfo {pages} {761} (\bibinfo {year} {2020})}\BibitemShut {NoStop}%
\bibitem [{\citenamefont {Helbig}\ \emph {et~al.}(2020)\citenamefont {Helbig}, \citenamefont {Hofmann}, \citenamefont {Imhof}, \citenamefont {Abdelghany}, \citenamefont {Kiessling}, \citenamefont {Molenkamp}, \citenamefont {Lee}, \citenamefont {Szameit}, \citenamefont {Greiter},\ and\ \citenamefont {Thomale}}]{Helbig2020generalBBC}%
  \BibitemOpen
  \bibfield  {author} {\bibinfo {author} {\bibfnamefont {T.}~\bibnamefont {Helbig}}, \bibinfo {author} {\bibfnamefont {T.}~\bibnamefont {Hofmann}}, \bibinfo {author} {\bibfnamefont {S.}~\bibnamefont {Imhof}}, \bibinfo {author} {\bibfnamefont {M.}~\bibnamefont {Abdelghany}}, \bibinfo {author} {\bibfnamefont {T.}~\bibnamefont {Kiessling}}, \bibinfo {author} {\bibfnamefont {L.~W.}\ \bibnamefont {Molenkamp}}, \bibinfo {author} {\bibfnamefont {C.~H.}\ \bibnamefont {Lee}}, \bibinfo {author} {\bibfnamefont {A.}~\bibnamefont {Szameit}}, \bibinfo {author} {\bibfnamefont {M.}~\bibnamefont {Greiter}}, \ and\ \bibinfo {author} {\bibfnamefont {R.}~\bibnamefont {Thomale}},\ }\bibfield  {title} {\enquote {\bibinfo {title} {Generalized bulk--boundary correspondence in non-hermitian topolectrical circuits},}\ }\href {\doibase 10.1038/s41567-020-0922-9} {\bibfield  {journal} {\bibinfo  {journal} {Nature Physics}\ }\textbf {\bibinfo {volume} {16}},\ \bibinfo {pages} {747} (\bibinfo {year} {2020})}\BibitemShut {NoStop}%
\bibitem [{\citenamefont {Okuma}\ \emph {et~al.}(2020)\citenamefont {Okuma}, \citenamefont {Kawabata}, \citenamefont {Shiozaki},\ and\ \citenamefont {Sato}}]{Okuma2020SkinTopology}%
  \BibitemOpen
  \bibfield  {author} {\bibinfo {author} {\bibfnamefont {N.}~\bibnamefont {Okuma}}, \bibinfo {author} {\bibfnamefont {K.}~\bibnamefont {Kawabata}}, \bibinfo {author} {\bibfnamefont {K.}~\bibnamefont {Shiozaki}}, \ and\ \bibinfo {author} {\bibfnamefont {M.}~\bibnamefont {Sato}},\ }\bibfield  {title} {\enquote {\bibinfo {title} {Topological origin of non-hermitian skin effects},}\ }\href {\doibase 10.1103/PhysRevLett.124.086801} {\bibfield  {journal} {\bibinfo  {journal} {Phys. Rev. Lett.}\ }\textbf {\bibinfo {volume} {124}},\ \bibinfo {pages} {086801} (\bibinfo {year} {2020})}\BibitemShut {NoStop}%
\bibitem [{\citenamefont {Zhang}\ \emph {et~al.}(2020{\natexlab{a}})\citenamefont {Zhang}, \citenamefont {Yang},\ and\ \citenamefont {Fang}}]{Zhang2020SpectralWinding}%
  \BibitemOpen
  \bibfield  {author} {\bibinfo {author} {\bibfnamefont {K.}~\bibnamefont {Zhang}}, \bibinfo {author} {\bibfnamefont {Z.}~\bibnamefont {Yang}}, \ and\ \bibinfo {author} {\bibfnamefont {C.}~\bibnamefont {Fang}},\ }\bibfield  {title} {\enquote {\bibinfo {title} {Correspondence between winding numbers and skin modes in non-hermitian systems},}\ }\href {\doibase 10.1103/PhysRevLett.125.126402} {\bibfield  {journal} {\bibinfo  {journal} {Phys. Rev. Lett.}\ }\textbf {\bibinfo {volume} {125}},\ \bibinfo {pages} {126402} (\bibinfo {year} {2020}{\natexlab{a}})}\BibitemShut {NoStop}%
\bibitem [{\citenamefont {Lee}\ and\ \citenamefont {Thomale}(2019)}]{Lee2019NHSEanatomy}%
  \BibitemOpen
  \bibfield  {author} {\bibinfo {author} {\bibfnamefont {C.~H.}\ \bibnamefont {Lee}}\ and\ \bibinfo {author} {\bibfnamefont {R.}~\bibnamefont {Thomale}},\ }\bibfield  {title} {\enquote {\bibinfo {title} {Anatomy of skin modes and topology in non-hermitian systems},}\ }\href {\doibase 10.1103/PhysRevB.99.201103} {\bibfield  {journal} {\bibinfo  {journal} {Phys. Rev. B}\ }\textbf {\bibinfo {volume} {99}},\ \bibinfo {pages} {201103} (\bibinfo {year} {2019})}\BibitemShut {NoStop}%
\bibitem [{\citenamefont {Ashida}\ \emph {et~al.}(2020)\citenamefont {Ashida}, \citenamefont {Gong},\ and\ \citenamefont {Ueda}}]{Ashida2021Non}%
  \BibitemOpen
  \bibfield  {author} {\bibinfo {author} {\bibfnamefont {Y.}~\bibnamefont {Ashida}}, \bibinfo {author} {\bibfnamefont {Z.}~\bibnamefont {Gong}}, \ and\ \bibinfo {author} {\bibfnamefont {M.}~\bibnamefont {Ueda}},\ }\bibfield  {title} {\enquote {\bibinfo {title} {Non-hermitian physics},}\ }\href {\doibase 10.1080/00018732.2021.1876991} {\bibfield  {journal} {\bibinfo  {journal} {Adv. Phys.}\ }\textbf {\bibinfo {volume} {69}},\ \bibinfo {pages} {249} (\bibinfo {year} {2020})}\BibitemShut {NoStop}%
\bibitem [{\citenamefont {Bergholtz}\ \emph {et~al.}(2021)\citenamefont {Bergholtz}, \citenamefont {Budich},\ and\ \citenamefont {Kunst}}]{Bergholtz2021RMP}%
  \BibitemOpen
  \bibfield  {author} {\bibinfo {author} {\bibfnamefont {E.~J.}\ \bibnamefont {Bergholtz}}, \bibinfo {author} {\bibfnamefont {J.~C.}\ \bibnamefont {Budich}}, \ and\ \bibinfo {author} {\bibfnamefont {F.~K.}\ \bibnamefont {Kunst}},\ }\bibfield  {title} {\enquote {\bibinfo {title} {Exceptional topology of non-hermitian systems},}\ }\href {\doibase 10.1103/RevModPhys.93.015005} {\bibfield  {journal} {\bibinfo  {journal} {Rev. Mod. Phys.}\ }\textbf {\bibinfo {volume} {93}},\ \bibinfo {pages} {015005} (\bibinfo {year} {2021})}\BibitemShut {NoStop}%
\bibitem [{\citenamefont {Gohsrich}\ \emph {et~al.}()\citenamefont {Gohsrich}, \citenamefont {Banerjee},\ and\ \citenamefont {Kunst}}]{Gohsrich2024Perspective}%
  \BibitemOpen
  \bibfield  {author} {\bibinfo {author} {\bibfnamefont {J.~T.}\ \bibnamefont {Gohsrich}}, \bibinfo {author} {\bibfnamefont {A.}~\bibnamefont {Banerjee}}, \ and\ \bibinfo {author} {\bibfnamefont {F.~K.}\ \bibnamefont {Kunst}},\ }\bibfield  {title} {\enquote {\bibinfo {title} {The non-hermitian skin effect: A perspective},}\ }\href {https://arxiv.org/abs/2410.23845} {\bibinfo  {journal} {arXiv:2410.23845}\ }\BibitemShut {NoStop}%
\bibitem [{\citenamefont {Wang}\ \emph {et~al.}(2024)\citenamefont {Wang}, \citenamefont {Song},\ and\ \citenamefont {Wang}}]{Wang2024Amoeba}%
  \BibitemOpen
\bibfield  {journal} {  }\bibfield  {author} {\bibinfo {author} {\bibfnamefont {H.-Y.}\ \bibnamefont {Wang}}, \bibinfo {author} {\bibfnamefont {F.}~\bibnamefont {Song}}, \ and\ \bibinfo {author} {\bibfnamefont {Z.}~\bibnamefont {Wang}},\ }\bibfield  {title} {\enquote {\bibinfo {title} {Amoeba formulation of non-bloch band theory in arbitrary dimensions},}\ }\href {\doibase 10.1103/PhysRevX.14.021011} {\bibfield  {journal} {\bibinfo  {journal} {Phys. Rev. X}\ }\textbf {\bibinfo {volume} {14}},\ \bibinfo {pages} {021011} (\bibinfo {year} {2024})}\BibitemShut {NoStop}%
\bibitem [{\citenamefont {Kawabata}\ and\ \citenamefont {Ryu}(2026)}]{kawabata2026nonhermitiandisorderedsystems}%
  \BibitemOpen
  \bibfield  {author} {\bibinfo {author} {\bibfnamefont {K.}~\bibnamefont {Kawabata}}\ and\ \bibinfo {author} {\bibfnamefont {S.}~\bibnamefont {Ryu}},\ }\href {https://arxiv.org/abs/2603.20393} {\enquote {\bibinfo {title} {Non-hermitian disordered systems},}\ } (\bibinfo {year} {2026}),\ \Eprint {http://arxiv.org/abs/2603.20393} {arXiv:2603.20393 [cond-mat.mes-hall]} \BibitemShut {NoStop}%
\bibitem [{\citenamefont {Zhang}\ \emph {et~al.}(2020{\natexlab{b}})\citenamefont {Zhang}, \citenamefont {Tang}, \citenamefont {Lang}, \citenamefont {Yan},\ and\ \citenamefont {Zhu}}]{Zhang2020NHtopoAnderson}%
  \BibitemOpen
  \bibfield  {author} {\bibinfo {author} {\bibfnamefont {D.-W.}\ \bibnamefont {Zhang}}, \bibinfo {author} {\bibfnamefont {L.-Z.}\ \bibnamefont {Tang}}, \bibinfo {author} {\bibfnamefont {L.-J.}\ \bibnamefont {Lang}}, \bibinfo {author} {\bibfnamefont {H.}~\bibnamefont {Yan}}, \ and\ \bibinfo {author} {\bibfnamefont {S.-L.}\ \bibnamefont {Zhu}},\ }\bibfield  {title} {\enquote {\bibinfo {title} {Non-hermitian topological anderson insulators},}\ }\href {\doibase 10.1007/s11433-020-1521-9} {\bibfield  {journal} {\bibinfo  {journal} {Science China Physics, Mechanics {\&} Astronomy}\ }\textbf {\bibinfo {volume} {63}},\ \bibinfo {pages} {267062} (\bibinfo {year} {2020}{\natexlab{b}})}\BibitemShut {NoStop}%
\bibitem [{\citenamefont {Tang}\ \emph {et~al.}(2020)\citenamefont {Tang}, \citenamefont {Zhang}, \citenamefont {Zhang},\ and\ \citenamefont {Zhang}}]{Tang2020TopoAnderson}%
  \BibitemOpen
  \bibfield  {author} {\bibinfo {author} {\bibfnamefont {L.-Z.}\ \bibnamefont {Tang}}, \bibinfo {author} {\bibfnamefont {L.-F.}\ \bibnamefont {Zhang}}, \bibinfo {author} {\bibfnamefont {G.-Q.}\ \bibnamefont {Zhang}}, \ and\ \bibinfo {author} {\bibfnamefont {D.-W.}\ \bibnamefont {Zhang}},\ }\bibfield  {title} {\enquote {\bibinfo {title} {Topological anderson insulators in two-dimensional non-hermitian disordered systems},}\ }\href {\doibase 10.1103/PhysRevA.101.063612} {\bibfield  {journal} {\bibinfo  {journal} {Phys. Rev. A}\ }\textbf {\bibinfo {volume} {101}},\ \bibinfo {pages} {063612} (\bibinfo {year} {2020})}\BibitemShut {NoStop}%
\bibitem [{\citenamefont {Liu}\ \emph {et~al.}(2021{\natexlab{a}})\citenamefont {Liu}, \citenamefont {Zhou}, \citenamefont {Wu}, \citenamefont {Zhang},\ and\ \citenamefont {Jiang}}]{Liu2021NHtopoAnderson2D}%
  \BibitemOpen
  \bibfield  {author} {\bibinfo {author} {\bibfnamefont {H.}~\bibnamefont {Liu}}, \bibinfo {author} {\bibfnamefont {J.-K.}\ \bibnamefont {Zhou}}, \bibinfo {author} {\bibfnamefont {B.-L.}\ \bibnamefont {Wu}}, \bibinfo {author} {\bibfnamefont {Z.-Q.}\ \bibnamefont {Zhang}}, \ and\ \bibinfo {author} {\bibfnamefont {H.}~\bibnamefont {Jiang}},\ }\bibfield  {title} {\enquote {\bibinfo {title} {Real-space topological invariant and higher-order topological anderson insulator in two-dimensional non-hermitian systems},}\ }\href {\doibase 10.1103/PhysRevB.103.224203} {\bibfield  {journal} {\bibinfo  {journal} {Phys. Rev. B}\ }\textbf {\bibinfo {volume} {103}},\ \bibinfo {pages} {224203} (\bibinfo {year} {2021}{\natexlab{a}})}\BibitemShut {NoStop}%
\bibitem [{\citenamefont {Lin}\ \emph {et~al.}(2022)\citenamefont {Lin}, \citenamefont {Li}, \citenamefont {Xiao}, \citenamefont {Wang}, \citenamefont {Yi},\ and\ \citenamefont {Xue}}]{Lin2022experimentNHtopoAI}%
  \BibitemOpen
  \bibfield  {author} {\bibinfo {author} {\bibfnamefont {Q.}~\bibnamefont {Lin}}, \bibinfo {author} {\bibfnamefont {T.}~\bibnamefont {Li}}, \bibinfo {author} {\bibfnamefont {L.}~\bibnamefont {Xiao}}, \bibinfo {author} {\bibfnamefont {K.}~\bibnamefont {Wang}}, \bibinfo {author} {\bibfnamefont {W.}~\bibnamefont {Yi}}, \ and\ \bibinfo {author} {\bibfnamefont {P.}~\bibnamefont {Xue}},\ }\bibfield  {title} {\enquote {\bibinfo {title} {Observation of non-hermitian topological anderson insulator in quantum dynamics},}\ }\href {\doibase 10.1038/s41467-022-30938-9} {\bibfield  {journal} {\bibinfo  {journal} {Nature Communications}\ }\textbf {\bibinfo {volume} {13}},\ \bibinfo {pages} {3229} (\bibinfo {year} {2022})}\BibitemShut {NoStop}%
\bibitem [{\citenamefont {Hamazaki}\ \emph {et~al.}(2019)\citenamefont {Hamazaki}, \citenamefont {Kawabata},\ and\ \citenamefont {Ueda}}]{Hamazaki2019nonHermitianMBL}%
  \BibitemOpen
  \bibfield  {author} {\bibinfo {author} {\bibfnamefont {R.}~\bibnamefont {Hamazaki}}, \bibinfo {author} {\bibfnamefont {K.}~\bibnamefont {Kawabata}}, \ and\ \bibinfo {author} {\bibfnamefont {M.}~\bibnamefont {Ueda}},\ }\bibfield  {title} {\enquote {\bibinfo {title} {Non-hermitian many-body localization},}\ }\href {\doibase 10.1103/PhysRevLett.123.090603} {\bibfield  {journal} {\bibinfo  {journal} {Phys. Rev. Lett.}\ }\textbf {\bibinfo {volume} {123}},\ \bibinfo {pages} {090603} (\bibinfo {year} {2019})}\BibitemShut {NoStop}%
\bibitem [{\citenamefont {Zhai}\ \emph {et~al.}(2020)\citenamefont {Zhai}, \citenamefont {Yin},\ and\ \citenamefont {Huang}}]{Zhai2020MBLquasiperiodic}%
  \BibitemOpen
  \bibfield  {author} {\bibinfo {author} {\bibfnamefont {L.-J.}\ \bibnamefont {Zhai}}, \bibinfo {author} {\bibfnamefont {S.}~\bibnamefont {Yin}}, \ and\ \bibinfo {author} {\bibfnamefont {G.-Y.}\ \bibnamefont {Huang}},\ }\bibfield  {title} {\enquote {\bibinfo {title} {Many-body localization in a non-hermitian quasiperiodic system},}\ }\href {\doibase 10.1103/PhysRevB.102.064206} {\bibfield  {journal} {\bibinfo  {journal} {Phys. Rev. B}\ }\textbf {\bibinfo {volume} {102}},\ \bibinfo {pages} {064206} (\bibinfo {year} {2020})}\BibitemShut {NoStop}%
\bibitem [{\citenamefont {Suthar}\ \emph {et~al.}(2022)\citenamefont {Suthar}, \citenamefont {Wang}, \citenamefont {Huang}, \citenamefont {Jen},\ and\ \citenamefont {You}}]{Suthar2022nhMBL}%
  \BibitemOpen
  \bibfield  {author} {\bibinfo {author} {\bibfnamefont {K.}~\bibnamefont {Suthar}}, \bibinfo {author} {\bibfnamefont {Y.-C.}\ \bibnamefont {Wang}}, \bibinfo {author} {\bibfnamefont {Y.-P.}\ \bibnamefont {Huang}}, \bibinfo {author} {\bibfnamefont {H.~H.}\ \bibnamefont {Jen}}, \ and\ \bibinfo {author} {\bibfnamefont {J.-S.}\ \bibnamefont {You}},\ }\bibfield  {title} {\enquote {\bibinfo {title} {Non-hermitian many-body localization with open boundaries},}\ }\href {\doibase 10.1103/PhysRevB.106.064208} {\bibfield  {journal} {\bibinfo  {journal} {Phys. Rev. B}\ }\textbf {\bibinfo {volume} {106}},\ \bibinfo {pages} {064208} (\bibinfo {year} {2022})}\BibitemShut {NoStop}%
\bibitem [{\citenamefont {Wang}\ \emph {et~al.}(2023)\citenamefont {Wang}, \citenamefont {Suthar}, \citenamefont {Jen}, \citenamefont {Hsu},\ and\ \citenamefont {You}}]{Wang2023skinMBL}%
  \BibitemOpen
  \bibfield  {author} {\bibinfo {author} {\bibfnamefont {Y.-C.}\ \bibnamefont {Wang}}, \bibinfo {author} {\bibfnamefont {K.}~\bibnamefont {Suthar}}, \bibinfo {author} {\bibfnamefont {H.~H.}\ \bibnamefont {Jen}}, \bibinfo {author} {\bibfnamefont {Y.-T.}\ \bibnamefont {Hsu}}, \ and\ \bibinfo {author} {\bibfnamefont {J.-S.}\ \bibnamefont {You}},\ }\bibfield  {title} {\enquote {\bibinfo {title} {Non-hermitian skin effects on thermal and many-body localized phases},}\ }\href {\doibase 10.1103/PhysRevB.107.L220205} {\bibfield  {journal} {\bibinfo  {journal} {Phys. Rev. B}\ }\textbf {\bibinfo {volume} {107}},\ \bibinfo {pages} {L220205} (\bibinfo {year} {2023})}\BibitemShut {NoStop}%
\bibitem [{\citenamefont {Roccati}\ \emph {et~al.}(2024)\citenamefont {Roccati}, \citenamefont {Balducci}, \citenamefont {Shir},\ and\ \citenamefont {Chenu}}]{Roccati2024nonHermitianMBLchaos}%
  \BibitemOpen
  \bibfield  {author} {\bibinfo {author} {\bibfnamefont {F.}~\bibnamefont {Roccati}}, \bibinfo {author} {\bibfnamefont {F.}~\bibnamefont {Balducci}}, \bibinfo {author} {\bibfnamefont {R.}~\bibnamefont {Shir}}, \ and\ \bibinfo {author} {\bibfnamefont {A.}~\bibnamefont {Chenu}},\ }\bibfield  {title} {\enquote {\bibinfo {title} {Diagnosing non-hermitian many-body localization and quantum chaos via singular value decomposition},}\ }\href {\doibase 10.1103/PhysRevB.109.L140201} {\bibfield  {journal} {\bibinfo  {journal} {Phys. Rev. B}\ }\textbf {\bibinfo {volume} {109}},\ \bibinfo {pages} {L140201} (\bibinfo {year} {2024})}\BibitemShut {NoStop}%
\bibitem [{\citenamefont {Marchetti}\ and\ \citenamefont {Simons}(2001)}]{Marchetti2001susyDOS}%
  \BibitemOpen
  \bibfield  {author} {\bibinfo {author} {\bibfnamefont {F.~M.}\ \bibnamefont {Marchetti}}\ and\ \bibinfo {author} {\bibfnamefont {B.~D.}\ \bibnamefont {Simons}},\ }\bibfield  {title} {\enquote {\bibinfo {title} {Optimal fluctuations and tail states of non-hermitian operators},}\ }\href {\doibase 10.1088/0305-4470/34/49/305} {\bibfield  {journal} {\bibinfo  {journal} {Journal of Physics A: Mathematical and General}\ }\textbf {\bibinfo {volume} {34}},\ \bibinfo {pages} {10805} (\bibinfo {year} {2001})}\BibitemShut {NoStop}%
\bibitem [{\citenamefont {Silvestrov}(2001)}]{Silvestrov2001NHtailDOS}%
  \BibitemOpen
  \bibfield  {author} {\bibinfo {author} {\bibfnamefont {P.~G.}\ \bibnamefont {Silvestrov}},\ }\bibfield  {title} {\enquote {\bibinfo {title} {Extended tail states in an imaginary random potential},}\ }\href {\doibase 10.1103/PhysRevB.64.075114} {\bibfield  {journal} {\bibinfo  {journal} {Phys. Rev. B}\ }\textbf {\bibinfo {volume} {64}},\ \bibinfo {pages} {075114} (\bibinfo {year} {2001})}\BibitemShut {NoStop}%
\bibitem [{\citenamefont {Longhi}(2025{\natexlab{a}})}]{Longhi2025LifshitzTail}%
  \BibitemOpen
  \bibfield  {author} {\bibinfo {author} {\bibfnamefont {S.}~\bibnamefont {Longhi}},\ }\bibfield  {title} {\enquote {\bibinfo {title} {Lifshitz tail states in non-hermitian disordered photonic lattices},}\ }\href {\doibase 10.1364/OL.551954} {\bibfield  {journal} {\bibinfo  {journal} {Opt. Lett.}\ }\textbf {\bibinfo {volume} {50}},\ \bibinfo {pages} {746} (\bibinfo {year} {2025}{\natexlab{a}})}\BibitemShut {NoStop}%
\bibitem [{\citenamefont {Hamazaki}\ \emph {et~al.}(2020)\citenamefont {Hamazaki}, \citenamefont {Kawabata}, \citenamefont {Kura},\ and\ \citenamefont {Ueda}}]{Hamazaki2020NHmatrixuniversality}%
  \BibitemOpen
  \bibfield  {author} {\bibinfo {author} {\bibfnamefont {R.}~\bibnamefont {Hamazaki}}, \bibinfo {author} {\bibfnamefont {K.}~\bibnamefont {Kawabata}}, \bibinfo {author} {\bibfnamefont {N.}~\bibnamefont {Kura}}, \ and\ \bibinfo {author} {\bibfnamefont {M.}~\bibnamefont {Ueda}},\ }\bibfield  {title} {\enquote {\bibinfo {title} {Universality classes of non-hermitian random matrices},}\ }\href {\doibase 10.1103/PhysRevResearch.2.023286} {\bibfield  {journal} {\bibinfo  {journal} {Phys. Rev. Res.}\ }\textbf {\bibinfo {volume} {2}},\ \bibinfo {pages} {023286} (\bibinfo {year} {2020})}\BibitemShut {NoStop}%
\bibitem [{\citenamefont {Chen}\ \emph {et~al.}(2025)\citenamefont {Chen}, \citenamefont {Kawabata}, \citenamefont {Kulkarni},\ and\ \citenamefont {Ryu}}]{Chen2025NHsigmaModel}%
  \BibitemOpen
  \bibfield  {author} {\bibinfo {author} {\bibfnamefont {Z.}~\bibnamefont {Chen}}, \bibinfo {author} {\bibfnamefont {K.}~\bibnamefont {Kawabata}}, \bibinfo {author} {\bibfnamefont {A.}~\bibnamefont {Kulkarni}}, \ and\ \bibinfo {author} {\bibfnamefont {S.}~\bibnamefont {Ryu}},\ }\bibfield  {title} {\enquote {\bibinfo {title} {Field theory of non-hermitian disordered systems},}\ }\href {\doibase 10.1103/PhysRevB.111.054203} {\bibfield  {journal} {\bibinfo  {journal} {Phys. Rev. B}\ }\textbf {\bibinfo {volume} {111}},\ \bibinfo {pages} {054203} (\bibinfo {year} {2025})}\BibitemShut {NoStop}%
\bibitem [{\citenamefont {Balasubrahmaniyam}\ \emph {et~al.}(2020)\citenamefont {Balasubrahmaniyam}, \citenamefont {Mondal},\ and\ \citenamefont {Mujumdar}}]{Balasubrahmaniyam2020Necklace}%
  \BibitemOpen
  \bibfield  {author} {\bibinfo {author} {\bibfnamefont {M.}~\bibnamefont {Balasubrahmaniyam}}, \bibinfo {author} {\bibfnamefont {S.}~\bibnamefont {Mondal}}, \ and\ \bibinfo {author} {\bibfnamefont {S.}~\bibnamefont {Mujumdar}},\ }\bibfield  {title} {\enquote {\bibinfo {title} {Necklace-state-mediated anomalous enhancement of transport in anderson-localized non-hermitian hybrid systems},}\ }\href {\doibase 10.1103/PhysRevLett.124.123901} {\bibfield  {journal} {\bibinfo  {journal} {Phys. Rev. Lett.}\ }\textbf {\bibinfo {volume} {124}},\ \bibinfo {pages} {123901} (\bibinfo {year} {2020})}\BibitemShut {NoStop}%
\bibitem [{\citenamefont {Weidemann}\ \emph {et~al.}(2021)\citenamefont {Weidemann}, \citenamefont {Kremer}, \citenamefont {Longhi},\ and\ \citenamefont {Szameit}}]{Weidemann2021NHtransport}%
  \BibitemOpen
  \bibfield  {author} {\bibinfo {author} {\bibfnamefont {S.}~\bibnamefont {Weidemann}}, \bibinfo {author} {\bibfnamefont {M.}~\bibnamefont {Kremer}}, \bibinfo {author} {\bibfnamefont {S.}~\bibnamefont {Longhi}}, \ and\ \bibinfo {author} {\bibfnamefont {A.}~\bibnamefont {Szameit}},\ }\bibfield  {title} {\enquote {\bibinfo {title} {Coexistence of dynamical delocalization and spectral localization through stochastic dissipation},}\ }\href {\doibase 10.1038/s41566-021-00823-w} {\bibfield  {journal} {\bibinfo  {journal} {Nature Photonics}\ }\textbf {\bibinfo {volume} {15}},\ \bibinfo {pages} {576} (\bibinfo {year} {2021})}\BibitemShut {NoStop}%
\bibitem [{\citenamefont {Longhi}(2023)}]{Longhi2023LocLindblad}%
  \BibitemOpen
  \bibfield  {author} {\bibinfo {author} {\bibfnamefont {S.}~\bibnamefont {Longhi}},\ }\bibfield  {title} {\enquote {\bibinfo {title} {Anderson localization in dissipative lattices},}\ }\href {\doibase https://doi.org/10.1002/andp.202200658} {\bibfield  {journal} {\bibinfo  {journal} {Annalen der Physik}\ }\textbf {\bibinfo {volume} {535}},\ \bibinfo {pages} {2200658} (\bibinfo {year} {2023})}\BibitemShut {NoStop}%
\bibitem [{\citenamefont {Yusipov}\ \emph {et~al.}(2018)\citenamefont {Yusipov}, \citenamefont {Laptyeva},\ and\ \citenamefont {Ivanchenko}}]{Yusipov2018AndersonJump}%
  \BibitemOpen
  \bibfield  {author} {\bibinfo {author} {\bibfnamefont {I.~I.}\ \bibnamefont {Yusipov}}, \bibinfo {author} {\bibfnamefont {T.~V.}\ \bibnamefont {Laptyeva}}, \ and\ \bibinfo {author} {\bibfnamefont {M.~V.}\ \bibnamefont {Ivanchenko}},\ }\bibfield  {title} {\enquote {\bibinfo {title} {Quantum jumps on anderson attractors},}\ }\href {\doibase 10.1103/PhysRevB.97.020301} {\bibfield  {journal} {\bibinfo  {journal} {Phys. Rev. B}\ }\textbf {\bibinfo {volume} {97}},\ \bibinfo {pages} {020301} (\bibinfo {year} {2018})}\BibitemShut {NoStop}%
\bibitem [{\citenamefont {Tzortzakakis}\ \emph {et~al.}(2021)\citenamefont {Tzortzakakis}, \citenamefont {Makris}, \citenamefont {Szameit},\ and\ \citenamefont {Economou}}]{Tzortzakakis2021NHtransport}%
  \BibitemOpen
  \bibfield  {author} {\bibinfo {author} {\bibfnamefont {A.~F.}\ \bibnamefont {Tzortzakakis}}, \bibinfo {author} {\bibfnamefont {K.~G.}\ \bibnamefont {Makris}}, \bibinfo {author} {\bibfnamefont {A.}~\bibnamefont {Szameit}}, \ and\ \bibinfo {author} {\bibfnamefont {E.~N.}\ \bibnamefont {Economou}},\ }\bibfield  {title} {\enquote {\bibinfo {title} {Transport and spectral features in non-hermitian open systems},}\ }\href {\doibase 10.1103/PhysRevResearch.3.013208} {\bibfield  {journal} {\bibinfo  {journal} {Phys. Rev. Res.}\ }\textbf {\bibinfo {volume} {3}},\ \bibinfo {pages} {013208} (\bibinfo {year} {2021})}\BibitemShut {NoStop}%
\bibitem [{\citenamefont {Leventis}\ \emph {et~al.}(2022)\citenamefont {Leventis}, \citenamefont {Makris},\ and\ \citenamefont {Economou}}]{Leventis2022NHjump}%
  \BibitemOpen
  \bibfield  {author} {\bibinfo {author} {\bibfnamefont {A.}~\bibnamefont {Leventis}}, \bibinfo {author} {\bibfnamefont {K.~G.}\ \bibnamefont {Makris}}, \ and\ \bibinfo {author} {\bibfnamefont {E.~N.}\ \bibnamefont {Economou}},\ }\bibfield  {title} {\enquote {\bibinfo {title} {Non-hermitian jumps in disordered lattices},}\ }\href {\doibase 10.1103/PhysRevB.106.064205} {\bibfield  {journal} {\bibinfo  {journal} {Phys. Rev. B}\ }\textbf {\bibinfo {volume} {106}},\ \bibinfo {pages} {064205} (\bibinfo {year} {2022})}\BibitemShut {NoStop}%
\bibitem [{\citenamefont {Sahoo}\ \emph {et~al.}(2022)\citenamefont {Sahoo}, \citenamefont {Vijay},\ and\ \citenamefont {Mujumdar}}]{Sahoo2022NHAndersonTransport}%
  \BibitemOpen
  \bibfield  {author} {\bibinfo {author} {\bibfnamefont {H.}~\bibnamefont {Sahoo}}, \bibinfo {author} {\bibfnamefont {R.}~\bibnamefont {Vijay}}, \ and\ \bibinfo {author} {\bibfnamefont {S.}~\bibnamefont {Mujumdar}},\ }\bibfield  {title} {\enquote {\bibinfo {title} {Anomalous transport regime in a non-hermitian anderson-localized hybrid system},}\ }\href {\doibase 10.1103/PhysRevResearch.4.043081} {\bibfield  {journal} {\bibinfo  {journal} {Phys. Rev. Res.}\ }\textbf {\bibinfo {volume} {4}},\ \bibinfo {pages} {043081} (\bibinfo {year} {2022})}\BibitemShut {NoStop}%
\bibitem [{\citenamefont {Tzortzakakis}\ \emph {et~al.}(2020)\citenamefont {Tzortzakakis}, \citenamefont {Makris},\ and\ \citenamefont {Economou}}]{Tzortzakakis2020complexdisorder}%
  \BibitemOpen
  \bibfield  {author} {\bibinfo {author} {\bibfnamefont {A.~F.}\ \bibnamefont {Tzortzakakis}}, \bibinfo {author} {\bibfnamefont {K.~G.}\ \bibnamefont {Makris}}, \ and\ \bibinfo {author} {\bibfnamefont {E.~N.}\ \bibnamefont {Economou}},\ }\bibfield  {title} {\enquote {\bibinfo {title} {Non-hermitian disorder in two-dimensional optical lattices},}\ }\href@noop {} {\bibfield  {journal} {\bibinfo  {journal} {Phys. Rev. B}\ }\textbf {\bibinfo {volume} {101}},\ \bibinfo {pages} {014202} (\bibinfo {year} {2020})}\BibitemShut {NoStop}%
\bibitem [{\citenamefont {Yusipov}\ \emph {et~al.}(2017)\citenamefont {Yusipov}, \citenamefont {Laptyeva}, \citenamefont {Denisov},\ and\ \citenamefont {Ivanchenko}}]{Yusipov2017OpenSysLocalization}%
  \BibitemOpen
  \bibfield  {author} {\bibinfo {author} {\bibfnamefont {I.}~\bibnamefont {Yusipov}}, \bibinfo {author} {\bibfnamefont {T.}~\bibnamefont {Laptyeva}}, \bibinfo {author} {\bibfnamefont {S.}~\bibnamefont {Denisov}}, \ and\ \bibinfo {author} {\bibfnamefont {M.}~\bibnamefont {Ivanchenko}},\ }\bibfield  {title} {\enquote {\bibinfo {title} {Localization in open quantum systems},}\ }\href {\doibase 10.1103/PhysRevLett.118.070402} {\bibfield  {journal} {\bibinfo  {journal} {Phys. Rev. Lett.}\ }\textbf {\bibinfo {volume} {118}},\ \bibinfo {pages} {070402} (\bibinfo {year} {2017})}\BibitemShut {NoStop}%
\bibitem [{\citenamefont {Li}\ \emph {et~al.}(2025)\citenamefont {Li}, \citenamefont {Chen},\ and\ \citenamefont {Wang}}]{Li2025universaldynamics}%
  \BibitemOpen
  \bibfield  {author} {\bibinfo {author} {\bibfnamefont {B.}~\bibnamefont {Li}}, \bibinfo {author} {\bibfnamefont {C.}~\bibnamefont {Chen}}, \ and\ \bibinfo {author} {\bibfnamefont {Z.}~\bibnamefont {Wang}},\ }\bibfield  {title} {\enquote {\bibinfo {title} {Universal non-hermitian transport in disordered systems},}\ }\href {\doibase 10.1103/z9m1-3mwb} {\bibfield  {journal} {\bibinfo  {journal} {Phys. Rev. Lett.}\ }\textbf {\bibinfo {volume} {135}},\ \bibinfo {pages} {033802} (\bibinfo {year} {2025})}\BibitemShut {NoStop}%
\bibitem [{\citenamefont {Xing}\ \emph {et~al.}(2025)\citenamefont {Xing}, \citenamefont {Chen},\ and\ \citenamefont {Hu}}]{Xing2025spreading}%
  \BibitemOpen
  \bibfield  {author} {\bibinfo {author} {\bibfnamefont {Z.-Y.}\ \bibnamefont {Xing}}, \bibinfo {author} {\bibfnamefont {S.}~\bibnamefont {Chen}}, \ and\ \bibinfo {author} {\bibfnamefont {H.}~\bibnamefont {Hu}},\ }\bibfield  {title} {\enquote {\bibinfo {title} {Universal spreading dynamics in quasiperiodic non-hermitian systems},}\ }\href {\doibase 10.1103/PhysRevB.111.L180203} {\bibfield  {journal} {\bibinfo  {journal} {Phys. Rev. B}\ }\textbf {\bibinfo {volume} {111}},\ \bibinfo {pages} {L180203} (\bibinfo {year} {2025})}\BibitemShut {NoStop}%
\bibitem [{\citenamefont {Shang}\ and\ \citenamefont {Hu}(2025{\natexlab{a}})}]{Shang2025spreading}%
  \BibitemOpen
  \bibfield  {author} {\bibinfo {author} {\bibfnamefont {J.}~\bibnamefont {Shang}}\ and\ \bibinfo {author} {\bibfnamefont {H.}~\bibnamefont {Hu}},\ }\bibfield  {title} {\enquote {\bibinfo {title} {Spreading dynamics in the hatano-nelson model},}\ }\href {\doibase 10.1103/wsmq-kmq9} {\bibfield  {journal} {\bibinfo  {journal} {Phys. Rev. B}\ }\textbf {\bibinfo {volume} {112}},\ \bibinfo {pages} {014205} (\bibinfo {year} {2025}{\natexlab{a}})}\BibitemShut {NoStop}%
\bibitem [{\citenamefont {Li}\ \emph {et~al.}(2024)\citenamefont {Li}, \citenamefont {Wang}, \citenamefont {Song},\ and\ \citenamefont {Wang}}]{Li2024nonblochdynamics}%
  \BibitemOpen
  \bibfield  {author} {\bibinfo {author} {\bibfnamefont {B.}~\bibnamefont {Li}}, \bibinfo {author} {\bibfnamefont {H.-R.}\ \bibnamefont {Wang}}, \bibinfo {author} {\bibfnamefont {F.}~\bibnamefont {Song}}, \ and\ \bibinfo {author} {\bibfnamefont {Z.}~\bibnamefont {Wang}},\ }\bibfield  {title} {\enquote {\bibinfo {title} {Non-bloch dynamics and topology in a classical nonequilibrium process},}\ }\href {\doibase 10.1103/PhysRevB.109.L201121} {\bibfield  {journal} {\bibinfo  {journal} {Phys. Rev. B}\ }\textbf {\bibinfo {volume} {109}},\ \bibinfo {pages} {L201121} (\bibinfo {year} {2024})}\BibitemShut {NoStop}%
\bibitem [{\citenamefont {Hatano}\ and\ \citenamefont {Nelson}(1996)}]{Hatano1996HNmodel}%
  \BibitemOpen
  \bibfield  {author} {\bibinfo {author} {\bibfnamefont {N.}~\bibnamefont {Hatano}}\ and\ \bibinfo {author} {\bibfnamefont {D.~R.}\ \bibnamefont {Nelson}},\ }\bibfield  {title} {\enquote {\bibinfo {title} {Localization transitions in non-hermitian quantum mechanics},}\ }\href {\doibase 10.1103/PhysRevLett.77.570} {\bibfield  {journal} {\bibinfo  {journal} {Phys. Rev. Lett.}\ }\textbf {\bibinfo {volume} {77}},\ \bibinfo {pages} {570} (\bibinfo {year} {1996})}\BibitemShut {NoStop}%
\bibitem [{\citenamefont {Hatano}\ and\ \citenamefont {Nelson}(1997)}]{Hatano1997HNmodel}%
  \BibitemOpen
  \bibfield  {author} {\bibinfo {author} {\bibfnamefont {N.}~\bibnamefont {Hatano}}\ and\ \bibinfo {author} {\bibfnamefont {D.~R.}\ \bibnamefont {Nelson}},\ }\bibfield  {title} {\enquote {\bibinfo {title} {Vortex pinning and non-hermitian quantum mechanics},}\ }\href {\doibase 10.1103/PhysRevB.56.8651} {\bibfield  {journal} {\bibinfo  {journal} {Phys. Rev. B}\ }\textbf {\bibinfo {volume} {56}},\ \bibinfo {pages} {8651} (\bibinfo {year} {1997})}\BibitemShut {NoStop}%
\bibitem [{\citenamefont {Efetov}(1997{\natexlab{a}})}]{Efetov1997DirectedChaos}%
  \BibitemOpen
  \bibfield  {author} {\bibinfo {author} {\bibfnamefont {K.~B.}\ \bibnamefont {Efetov}},\ }\bibfield  {title} {\enquote {\bibinfo {title} {Directed quantum chaos},}\ }\href {\doibase 10.1103/PhysRevLett.79.491} {\bibfield  {journal} {\bibinfo  {journal} {Phys. Rev. Lett.}\ }\textbf {\bibinfo {volume} {79}},\ \bibinfo {pages} {491} (\bibinfo {year} {1997}{\natexlab{a}})}\BibitemShut {NoStop}%
\bibitem [{\citenamefont {Efetov}(1997{\natexlab{b}})}]{Efetov1997DirectedDisorder}%
  \BibitemOpen
  \bibfield  {author} {\bibinfo {author} {\bibfnamefont {K.~B.}\ \bibnamefont {Efetov}},\ }\bibfield  {title} {\enquote {\bibinfo {title} {Quantum disordered systems with a direction},}\ }\href {\doibase 10.1103/PhysRevB.56.9630} {\bibfield  {journal} {\bibinfo  {journal} {Phys. Rev. B}\ }\textbf {\bibinfo {volume} {56}},\ \bibinfo {pages} {9630} (\bibinfo {year} {1997}{\natexlab{b}})}\BibitemShut {NoStop}%
\bibitem [{\citenamefont {Brouwer}\ \emph {et~al.}(1997)\citenamefont {Brouwer}, \citenamefont {Silvestrov},\ and\ \citenamefont {Beenakker}}]{Brouwer1997DirectedLocalization}%
  \BibitemOpen
  \bibfield  {author} {\bibinfo {author} {\bibfnamefont {P.~W.}\ \bibnamefont {Brouwer}}, \bibinfo {author} {\bibfnamefont {P.~G.}\ \bibnamefont {Silvestrov}}, \ and\ \bibinfo {author} {\bibfnamefont {C.~W.~J.}\ \bibnamefont {Beenakker}},\ }\bibfield  {title} {\enquote {\bibinfo {title} {Theory of directed localization in one dimension},}\ }\href {\doibase 10.1103/PhysRevB.56.R4333} {\bibfield  {journal} {\bibinfo  {journal} {Phys. Rev. B}\ }\textbf {\bibinfo {volume} {56}},\ \bibinfo {pages} {R4333} (\bibinfo {year} {1997})}\BibitemShut {NoStop}%
\bibitem [{\citenamefont {Hatano}\ and\ \citenamefont {Nelson}(1998{\natexlab{a}})}]{Hatano1998delocalization}%
  \BibitemOpen
  \bibfield  {author} {\bibinfo {author} {\bibfnamefont {N.}~\bibnamefont {Hatano}}\ and\ \bibinfo {author} {\bibfnamefont {D.~R.}\ \bibnamefont {Nelson}},\ }\bibfield  {title} {\enquote {\bibinfo {title} {Non-hermitian delocalization and eigenfunctions},}\ }\href {\doibase 10.1103/PhysRevB.58.8384} {\bibfield  {journal} {\bibinfo  {journal} {Phys. Rev. B}\ }\textbf {\bibinfo {volume} {58}},\ \bibinfo {pages} {8384} (\bibinfo {year} {1998}{\natexlab{a}})}\BibitemShut {NoStop}%
\bibitem [{\citenamefont {Nelson}\ and\ \citenamefont {Shnerb}(1998)}]{Nelson1998NHbiology}%
  \BibitemOpen
  \bibfield  {author} {\bibinfo {author} {\bibfnamefont {D.~R.}\ \bibnamefont {Nelson}}\ and\ \bibinfo {author} {\bibfnamefont {N.~M.}\ \bibnamefont {Shnerb}},\ }\bibfield  {title} {\enquote {\bibinfo {title} {Non-hermitian localization and population biology},}\ }\href {\doibase 10.1103/PhysRevE.58.1383} {\bibfield  {journal} {\bibinfo  {journal} {Phys. Rev. E}\ }\textbf {\bibinfo {volume} {58}},\ \bibinfo {pages} {1383} (\bibinfo {year} {1998})}\BibitemShut {NoStop}%
\bibitem [{\citenamefont {Shnerb}\ and\ \citenamefont {Nelson}(1998)}]{Shnerb1998WindingNumber}%
  \BibitemOpen
  \bibfield  {author} {\bibinfo {author} {\bibfnamefont {N.~M.}\ \bibnamefont {Shnerb}}\ and\ \bibinfo {author} {\bibfnamefont {D.~R.}\ \bibnamefont {Nelson}},\ }\bibfield  {title} {\enquote {\bibinfo {title} {Winding numbers, complex currents, and non-hermitian localization},}\ }\href {\doibase 10.1103/PhysRevLett.80.5172} {\bibfield  {journal} {\bibinfo  {journal} {Phys. Rev. Lett.}\ }\textbf {\bibinfo {volume} {80}},\ \bibinfo {pages} {5172} (\bibinfo {year} {1998})}\BibitemShut {NoStop}%
\bibitem [{\citenamefont {Goldsheid}\ and\ \citenamefont {Khoruzhenko}(1998)}]{Goldsheid1998NHAnderson}%
  \BibitemOpen
  \bibfield  {author} {\bibinfo {author} {\bibfnamefont {I.~Y.}\ \bibnamefont {Goldsheid}}\ and\ \bibinfo {author} {\bibfnamefont {B.~A.}\ \bibnamefont {Khoruzhenko}},\ }\bibfield  {title} {\enquote {\bibinfo {title} {Distribution of eigenvalues in non-hermitian anderson models},}\ }\href {\doibase 10.1103/PhysRevLett.80.2897} {\bibfield  {journal} {\bibinfo  {journal} {Phys. Rev. Lett.}\ }\textbf {\bibinfo {volume} {80}},\ \bibinfo {pages} {2897} (\bibinfo {year} {1998})}\BibitemShut {NoStop}%
\bibitem [{\citenamefont {Feinberg}\ and\ \citenamefont {Zee}(1999{\natexlab{a}})}]{Feinberg1999NHspectrum}%
  \BibitemOpen
  \bibfield  {author} {\bibinfo {author} {\bibfnamefont {J.}~\bibnamefont {Feinberg}}\ and\ \bibinfo {author} {\bibfnamefont {A.}~\bibnamefont {Zee}},\ }\bibfield  {title} {\enquote {\bibinfo {title} {Spectral curves of non-hermitian hamiltonians},}\ }\href {\doibase https://doi.org/10.1016/S0550-3213(99)00246-1} {\bibfield  {journal} {\bibinfo  {journal} {Nuclear Physics B}\ }\textbf {\bibinfo {volume} {552}},\ \bibinfo {pages} {599} (\bibinfo {year} {1999}{\natexlab{a}})}\BibitemShut {NoStop}%
\bibitem [{\citenamefont {Feinberg}\ and\ \citenamefont {Zee}(1999{\natexlab{b}})}]{Feinberg1999delocalization}%
  \BibitemOpen
  \bibfield  {author} {\bibinfo {author} {\bibfnamefont {J.}~\bibnamefont {Feinberg}}\ and\ \bibinfo {author} {\bibfnamefont {A.}~\bibnamefont {Zee}},\ }\bibfield  {title} {\enquote {\bibinfo {title} {Non-hermitian localization and delocalization},}\ }\href {\doibase 10.1103/PhysRevE.59.6433} {\bibfield  {journal} {\bibinfo  {journal} {Phys. Rev. E}\ }\textbf {\bibinfo {volume} {59}},\ \bibinfo {pages} {6433} (\bibinfo {year} {1999}{\natexlab{b}})}\BibitemShut {NoStop}%
\bibitem [{\citenamefont {Kolesnikov}\ and\ \citenamefont {Efetov}(2000)}]{Kolesnikov2000delocTransition}%
  \BibitemOpen
  \bibfield  {author} {\bibinfo {author} {\bibfnamefont {A.~V.}\ \bibnamefont {Kolesnikov}}\ and\ \bibinfo {author} {\bibfnamefont {K.~B.}\ \bibnamefont {Efetov}},\ }\bibfield  {title} {\enquote {\bibinfo {title} {Localization-delocalization transition in non-hermitian disordered systems},}\ }\href {\doibase 10.1103/PhysRevLett.84.5600} {\bibfield  {journal} {\bibinfo  {journal} {Phys. Rev. Lett.}\ }\textbf {\bibinfo {volume} {84}},\ \bibinfo {pages} {5600} (\bibinfo {year} {2000})}\BibitemShut {NoStop}%
\bibitem [{\citenamefont {Jiang}\ \emph {et~al.}(2019)\citenamefont {Jiang}, \citenamefont {Lang}, \citenamefont {Yang}, \citenamefont {Zhu},\ and\ \citenamefont {Chen}}]{Jiang2019NHSEquasiperiodic}%
  \BibitemOpen
  \bibfield  {author} {\bibinfo {author} {\bibfnamefont {H.}~\bibnamefont {Jiang}}, \bibinfo {author} {\bibfnamefont {L.-J.}\ \bibnamefont {Lang}}, \bibinfo {author} {\bibfnamefont {C.}~\bibnamefont {Yang}}, \bibinfo {author} {\bibfnamefont {S.-L.}\ \bibnamefont {Zhu}}, \ and\ \bibinfo {author} {\bibfnamefont {S.}~\bibnamefont {Chen}},\ }\bibfield  {title} {\enquote {\bibinfo {title} {Interplay of non-hermitian skin effects and anderson localization in nonreciprocal quasiperiodic lattices},}\ }\href {\doibase 10.1103/PhysRevB.100.054301} {\bibfield  {journal} {\bibinfo  {journal} {Phys. Rev. B}\ }\textbf {\bibinfo {volume} {100}},\ \bibinfo {pages} {054301} (\bibinfo {year} {2019})}\BibitemShut {NoStop}%
\bibitem [{\citenamefont {Longhi}(2019{\natexlab{a}})}]{Longhi2019TopoQuasicrystal}%
  \BibitemOpen
  \bibfield  {author} {\bibinfo {author} {\bibfnamefont {S.}~\bibnamefont {Longhi}},\ }\bibfield  {title} {\enquote {\bibinfo {title} {Topological phase transition in non-hermitian quasicrystals},}\ }\href {\doibase 10.1103/PhysRevLett.122.237601} {\bibfield  {journal} {\bibinfo  {journal} {Phys. Rev. Lett.}\ }\textbf {\bibinfo {volume} {122}},\ \bibinfo {pages} {237601} (\bibinfo {year} {2019}{\natexlab{a}})}\BibitemShut {NoStop}%
\bibitem [{\citenamefont {Longhi}(2019{\natexlab{b}})}]{Longhi2019NHandersionTransition}%
  \BibitemOpen
  \bibfield  {author} {\bibinfo {author} {\bibfnamefont {S.}~\bibnamefont {Longhi}},\ }\bibfield  {title} {\enquote {\bibinfo {title} {Metal-insulator phase transition in a non-hermitian aubry-andr\'e-harper model},}\ }\href {\doibase 10.1103/PhysRevB.100.125157} {\bibfield  {journal} {\bibinfo  {journal} {Phys. Rev. B}\ }\textbf {\bibinfo {volume} {100}},\ \bibinfo {pages} {125157} (\bibinfo {year} {2019}{\natexlab{b}})}\BibitemShut {NoStop}%
\bibitem [{\citenamefont {Huang}\ and\ \citenamefont {Shklovskii}(2020)}]{Huang2020nhAnderson}%
  \BibitemOpen
  \bibfield  {author} {\bibinfo {author} {\bibfnamefont {Y.}~\bibnamefont {Huang}}\ and\ \bibinfo {author} {\bibfnamefont {B.~I.}\ \bibnamefont {Shklovskii}},\ }\bibfield  {title} {\enquote {\bibinfo {title} {Anderson transition in three-dimensional systems with non-hermitian disorder},}\ }\href {\doibase 10.1103/PhysRevB.101.014204} {\bibfield  {journal} {\bibinfo  {journal} {Phys. Rev. B}\ }\textbf {\bibinfo {volume} {101}},\ \bibinfo {pages} {014204} (\bibinfo {year} {2020})}\BibitemShut {NoStop}%
\bibitem [{\citenamefont {Liu}\ \emph {et~al.}(2020)\citenamefont {Liu}, \citenamefont {Guo}, \citenamefont {Pu},\ and\ \citenamefont {Longhi}}]{Liu2020NHquasiperiodic}%
  \BibitemOpen
  \bibfield  {author} {\bibinfo {author} {\bibfnamefont {T.}~\bibnamefont {Liu}}, \bibinfo {author} {\bibfnamefont {H.}~\bibnamefont {Guo}}, \bibinfo {author} {\bibfnamefont {Y.}~\bibnamefont {Pu}}, \ and\ \bibinfo {author} {\bibfnamefont {S.}~\bibnamefont {Longhi}},\ }\bibfield  {title} {\enquote {\bibinfo {title} {Generalized aubry-andr\'e self-duality and mobility edges in non-hermitian quasiperiodic lattices},}\ }\href {\doibase 10.1103/PhysRevB.102.024205} {\bibfield  {journal} {\bibinfo  {journal} {Phys. Rev. B}\ }\textbf {\bibinfo {volume} {102}},\ \bibinfo {pages} {024205} (\bibinfo {year} {2020})}\BibitemShut {NoStop}%
\bibitem [{\citenamefont {Zeng}\ and\ \citenamefont {Xu}(2020)}]{Zeng2020NHmobility}%
  \BibitemOpen
  \bibfield  {author} {\bibinfo {author} {\bibfnamefont {Q.-B.}\ \bibnamefont {Zeng}}\ and\ \bibinfo {author} {\bibfnamefont {Y.}~\bibnamefont {Xu}},\ }\bibfield  {title} {\enquote {\bibinfo {title} {Winding numbers and generalized mobility edges in non-hermitian systems},}\ }\href {\doibase 10.1103/PhysRevResearch.2.033052} {\bibfield  {journal} {\bibinfo  {journal} {Phys. Rev. Res.}\ }\textbf {\bibinfo {volume} {2}},\ \bibinfo {pages} {033052} (\bibinfo {year} {2020})}\BibitemShut {NoStop}%
\bibitem [{\citenamefont {Kawabata}\ and\ \citenamefont {Ryu}(2021)}]{Kawabata2021NonunitaryScaling}%
  \BibitemOpen
  \bibfield  {author} {\bibinfo {author} {\bibfnamefont {K.}~\bibnamefont {Kawabata}}\ and\ \bibinfo {author} {\bibfnamefont {S.}~\bibnamefont {Ryu}},\ }\bibfield  {title} {\enquote {\bibinfo {title} {Nonunitary scaling theory of non-hermitian localization},}\ }\href {\doibase 10.1103/PhysRevLett.126.166801} {\bibfield  {journal} {\bibinfo  {journal} {Phys. Rev. Lett.}\ }\textbf {\bibinfo {volume} {126}},\ \bibinfo {pages} {166801} (\bibinfo {year} {2021})}\BibitemShut {NoStop}%
\bibitem [{\citenamefont {Weidemann}\ \emph {et~al.}(2022)\citenamefont {Weidemann}, \citenamefont {Kremer}, \citenamefont {Longhi},\ and\ \citenamefont {Szameit}}]{Weidemann2022NHfloquetquasicrystal}%
  \BibitemOpen
  \bibfield  {author} {\bibinfo {author} {\bibfnamefont {S.}~\bibnamefont {Weidemann}}, \bibinfo {author} {\bibfnamefont {M.}~\bibnamefont {Kremer}}, \bibinfo {author} {\bibfnamefont {S.}~\bibnamefont {Longhi}}, \ and\ \bibinfo {author} {\bibfnamefont {A.}~\bibnamefont {Szameit}},\ }\bibfield  {title} {\enquote {\bibinfo {title} {Topological triple phase transition in non-hermitian floquet quasicrystals},}\ }\href {\doibase 10.1038/s41586-021-04253-0} {\bibfield  {journal} {\bibinfo  {journal} {Nature}\ }\textbf {\bibinfo {volume} {601}},\ \bibinfo {pages} {354} (\bibinfo {year} {2022})}\BibitemShut {NoStop}%
\bibitem [{\citenamefont {Luo}\ \emph {et~al.}(2021{\natexlab{a}})\citenamefont {Luo}, \citenamefont {Ohtsuki},\ and\ \citenamefont {Shindou}}]{Luo2021transfermatrixAT}%
  \BibitemOpen
  \bibfield  {author} {\bibinfo {author} {\bibfnamefont {X.}~\bibnamefont {Luo}}, \bibinfo {author} {\bibfnamefont {T.}~\bibnamefont {Ohtsuki}}, \ and\ \bibinfo {author} {\bibfnamefont {R.}~\bibnamefont {Shindou}},\ }\bibfield  {title} {\enquote {\bibinfo {title} {Transfer matrix study of the anderson transition in non-hermitian systems},}\ }\href {\doibase 10.1103/PhysRevB.104.104203} {\bibfield  {journal} {\bibinfo  {journal} {Phys. Rev. B}\ }\textbf {\bibinfo {volume} {104}},\ \bibinfo {pages} {104203} (\bibinfo {year} {2021}{\natexlab{a}})}\BibitemShut {NoStop}%
\bibitem [{\citenamefont {Luo}\ \emph {et~al.}(2021{\natexlab{b}})\citenamefont {Luo}, \citenamefont {Ohtsuki},\ and\ \citenamefont {Shindou}}]{Luo2021NHandersonTransition}%
  \BibitemOpen
  \bibfield  {author} {\bibinfo {author} {\bibfnamefont {X.}~\bibnamefont {Luo}}, \bibinfo {author} {\bibfnamefont {T.}~\bibnamefont {Ohtsuki}}, \ and\ \bibinfo {author} {\bibfnamefont {R.}~\bibnamefont {Shindou}},\ }\bibfield  {title} {\enquote {\bibinfo {title} {Universality classes of the anderson transitions driven by non-hermitian disorder},}\ }\href {\doibase 10.1103/PhysRevLett.126.090402} {\bibfield  {journal} {\bibinfo  {journal} {Phys. Rev. Lett.}\ }\textbf {\bibinfo {volume} {126}},\ \bibinfo {pages} {090402} (\bibinfo {year} {2021}{\natexlab{b}})}\BibitemShut {NoStop}%
\bibitem [{\citenamefont {Liu}\ \emph {et~al.}(2021{\natexlab{b}})\citenamefont {Liu}, \citenamefont {Zhou},\ and\ \citenamefont {Chen}}]{Liu2021NHquasicrystal}%
  \BibitemOpen
  \bibfield  {author} {\bibinfo {author} {\bibfnamefont {Y.}~\bibnamefont {Liu}}, \bibinfo {author} {\bibfnamefont {Q.}~\bibnamefont {Zhou}}, \ and\ \bibinfo {author} {\bibfnamefont {S.}~\bibnamefont {Chen}},\ }\bibfield  {title} {\enquote {\bibinfo {title} {Localization transition, spectrum structure, and winding numbers for one-dimensional non-hermitian quasicrystals},}\ }\href {\doibase 10.1103/PhysRevB.104.024201} {\bibfield  {journal} {\bibinfo  {journal} {Phys. Rev. B}\ }\textbf {\bibinfo {volume} {104}},\ \bibinfo {pages} {024201} (\bibinfo {year} {2021}{\natexlab{b}})}\BibitemShut {NoStop}%
\bibitem [{\citenamefont {Luo}\ \emph {et~al.}(2022)\citenamefont {Luo}, \citenamefont {Xiao}, \citenamefont {Kawabata}, \citenamefont {Ohtsuki},\ and\ \citenamefont {Shindou}}]{Luo2022NHandersonTransition}%
  \BibitemOpen
  \bibfield  {author} {\bibinfo {author} {\bibfnamefont {X.}~\bibnamefont {Luo}}, \bibinfo {author} {\bibfnamefont {Z.}~\bibnamefont {Xiao}}, \bibinfo {author} {\bibfnamefont {K.}~\bibnamefont {Kawabata}}, \bibinfo {author} {\bibfnamefont {T.}~\bibnamefont {Ohtsuki}}, \ and\ \bibinfo {author} {\bibfnamefont {R.}~\bibnamefont {Shindou}},\ }\bibfield  {title} {\enquote {\bibinfo {title} {Unifying the anderson transitions in hermitian and non-hermitian systems},}\ }\href {\doibase 10.1103/PhysRevResearch.4.L022035} {\bibfield  {journal} {\bibinfo  {journal} {Phys. Rev. Res.}\ }\textbf {\bibinfo {volume} {4}},\ \bibinfo {pages} {L022035} (\bibinfo {year} {2022})}\BibitemShut {NoStop}%
\bibitem [{\citenamefont {Liu}\ \emph {et~al.}(2024)\citenamefont {Liu}, \citenamefont {Wang}, \citenamefont {Yang}, \citenamefont {Jie},\ and\ \citenamefont {Wang}}]{Liu2024dissipationLocal}%
  \BibitemOpen
  \bibfield  {author} {\bibinfo {author} {\bibfnamefont {Y.}~\bibnamefont {Liu}}, \bibinfo {author} {\bibfnamefont {Z.}~\bibnamefont {Wang}}, \bibinfo {author} {\bibfnamefont {C.}~\bibnamefont {Yang}}, \bibinfo {author} {\bibfnamefont {J.}~\bibnamefont {Jie}}, \ and\ \bibinfo {author} {\bibfnamefont {Y.}~\bibnamefont {Wang}},\ }\bibfield  {title} {\enquote {\bibinfo {title} {Dissipation-induced extended-localized transition},}\ }\href {\doibase 10.1103/PhysRevLett.132.216301} {\bibfield  {journal} {\bibinfo  {journal} {Phys. Rev. Lett.}\ }\textbf {\bibinfo {volume} {132}},\ \bibinfo {pages} {216301} (\bibinfo {year} {2024})}\BibitemShut {NoStop}%
\bibitem [{\citenamefont {Wang}\ \emph {et~al.}(2025{\natexlab{a}})\citenamefont {Wang}, \citenamefont {Wang},\ and\ \citenamefont {Ma}}]{Wang2025nonBlochAT}%
  \BibitemOpen
  \bibfield  {author} {\bibinfo {author} {\bibfnamefont {W.}~\bibnamefont {Wang}}, \bibinfo {author} {\bibfnamefont {X.}~\bibnamefont {Wang}}, \ and\ \bibinfo {author} {\bibfnamefont {G.}~\bibnamefont {Ma}},\ }\bibfield  {title} {\enquote {\bibinfo {title} {Anderson transition at complex energies in one-dimensional parity-time-symmetric disordered systems},}\ }\href {\doibase 10.1103/PhysRevLett.134.066301} {\bibfield  {journal} {\bibinfo  {journal} {Phys. Rev. Lett.}\ }\textbf {\bibinfo {volume} {134}},\ \bibinfo {pages} {066301} (\bibinfo {year} {2025}{\natexlab{a}})}\BibitemShut {NoStop}%
\bibitem [{\citenamefont {Longhi}(2025{\natexlab{b}})}]{Longhi2025erratic}%
  \BibitemOpen
  \bibfield  {author} {\bibinfo {author} {\bibfnamefont {S.}~\bibnamefont {Longhi}},\ }\bibfield  {title} {\enquote {\bibinfo {title} {Erratic non-hermitian skin localization},}\ }\href {\doibase 10.1103/PhysRevLett.134.196302} {\bibfield  {journal} {\bibinfo  {journal} {Phys. Rev. Lett.}\ }\textbf {\bibinfo {volume} {134}},\ \bibinfo {pages} {196302} (\bibinfo {year} {2025}{\natexlab{b}})}\BibitemShut {NoStop}%
\bibitem [{\citenamefont {Wang}\ \emph {et~al.}(2025{\natexlab{b}})\citenamefont {Wang}, \citenamefont {He}, \citenamefont {Wang},\ and\ \citenamefont {Ren}}]{Wang2025delocalization}%
  \BibitemOpen
  \bibfield  {author} {\bibinfo {author} {\bibfnamefont {C.}~\bibnamefont {Wang}}, \bibinfo {author} {\bibfnamefont {W.}~\bibnamefont {He}}, \bibinfo {author} {\bibfnamefont {X.~R.}\ \bibnamefont {Wang}}, \ and\ \bibinfo {author} {\bibfnamefont {H.}~\bibnamefont {Ren}},\ }\bibfield  {title} {\enquote {\bibinfo {title} {Unified one-parameter scaling function for anderson localization transitions in nonreciprocal non-hermitian systems},}\ }\href {\doibase 10.1103/PhysRevLett.134.176301} {\bibfield  {journal} {\bibinfo  {journal} {Phys. Rev. Lett.}\ }\textbf {\bibinfo {volume} {134}},\ \bibinfo {pages} {176301} (\bibinfo {year} {2025}{\natexlab{b}})}\BibitemShut {NoStop}%
\bibitem [{\citenamefont {Huang}\ \emph {et~al.}(2025)\citenamefont {Huang}, \citenamefont {Hu}, \citenamefont {Xue},\ and\ \citenamefont {Wang}}]{Huang2025UniversalScaling}%
  \BibitemOpen
  \bibfield  {author} {\bibinfo {author} {\bibfnamefont {Y.-Q.}\ \bibnamefont {Huang}}, \bibinfo {author} {\bibfnamefont {Y.-M.}\ \bibnamefont {Hu}}, \bibinfo {author} {\bibfnamefont {W.-T.}\ \bibnamefont {Xue}}, \ and\ \bibinfo {author} {\bibfnamefont {Z.}~\bibnamefont {Wang}},\ }\bibfield  {title} {\enquote {\bibinfo {title} {Universal scaling of green's functions in disordered non-hermitian systems},}\ }\href {\doibase 10.1103/PhysRevB.111.L060203} {\bibfield  {journal} {\bibinfo  {journal} {Phys. Rev. B}\ }\textbf {\bibinfo {volume} {111}},\ \bibinfo {pages} {L060203} (\bibinfo {year} {2025})}\BibitemShut {NoStop}%
\bibitem [{\citenamefont {Wang}\ \emph {et~al.}(2025{\natexlab{c}})\citenamefont {Wang}, \citenamefont {He}, \citenamefont {Wang},\ and\ \citenamefont {Ren}}]{Wang2025delocalization3D}%
  \BibitemOpen
  \bibfield  {author} {\bibinfo {author} {\bibfnamefont {C.}~\bibnamefont {Wang}}, \bibinfo {author} {\bibfnamefont {W.}~\bibnamefont {He}}, \bibinfo {author} {\bibfnamefont {X.~R.}\ \bibnamefont {Wang}}, \ and\ \bibinfo {author} {\bibfnamefont {H.}~\bibnamefont {Ren}},\ }\bibfield  {title} {\enquote {\bibinfo {title} {One-parameter scaling function for anderson localization transitions in high-dimensional lattices with nonreciprocity},}\ }\href {\doibase 10.1103/dvn4-2z9l} {\bibfield  {journal} {\bibinfo  {journal} {Phys. Rev. B}\ }\textbf {\bibinfo {volume} {112}},\ \bibinfo {pages} {214206} (\bibinfo {year} {2025}{\natexlab{c}})}\BibitemShut {NoStop}%
\bibitem [{\citenamefont {Shang}\ and\ \citenamefont {Hu}(2025{\natexlab{b}})}]{shang2025anisotropicLocalization}%
  \BibitemOpen
  \bibfield  {author} {\bibinfo {author} {\bibfnamefont {J.}~\bibnamefont {Shang}}\ and\ \bibinfo {author} {\bibfnamefont {H.}~\bibnamefont {Hu}},\ }\href {https://arxiv.org/abs/2507.14523} {\enquote {\bibinfo {title} {Anisotropic anderson localization in higher-dimensional nonreciprocal lattices},}\ } (\bibinfo {year} {2025}{\natexlab{b}}),\ \Eprint {http://arxiv.org/abs/2507.14523} {arXiv:2507.14523 [cond-mat.dis-nn]} \BibitemShut {NoStop}%
\bibitem [{\citenamefont {Wang}\ \emph {et~al.}(2026)\citenamefont {Wang}, \citenamefont {Wang},\ and\ \citenamefont {Ren}}]{wang2026skinanderson}%
  \BibitemOpen
  \bibfield  {author} {\bibinfo {author} {\bibfnamefont {C.}~\bibnamefont {Wang}}, \bibinfo {author} {\bibfnamefont {X.~R.}\ \bibnamefont {Wang}}, \ and\ \bibinfo {author} {\bibfnamefont {H.}~\bibnamefont {Ren}},\ }\href {https://arxiv.org/abs/2603.27069} {\enquote {\bibinfo {title} {Skin-anderson localization transitions in disordered hybrid-nonreciprocal systems},}\ } (\bibinfo {year} {2026}),\ \Eprint {http://arxiv.org/abs/2603.27069} {arXiv:2603.27069 [cond-mat.dis-nn]} \BibitemShut {NoStop}%
\bibitem [{\citenamefont {Longhi}(2021)}]{Longhi2021disorderNHSE}%
  \BibitemOpen
  \bibfield  {author} {\bibinfo {author} {\bibfnamefont {S.}~\bibnamefont {Longhi}},\ }\bibfield  {title} {\enquote {\bibinfo {title} {Spectral deformations in non-hermitian lattices with disorder and skin effect: A solvable model},}\ }\href {\doibase 10.1103/PhysRevB.103.144202} {\bibfield  {journal} {\bibinfo  {journal} {Phys. Rev. B}\ }\textbf {\bibinfo {volume} {103}},\ \bibinfo {pages} {144202} (\bibinfo {year} {2021})}\BibitemShut {NoStop}%
\bibitem [{\citenamefont {{Sun}}\ and\ \citenamefont {{Hu}}(2025)}]{Sun2025skinAndersonTransition}%
  \BibitemOpen
  \bibfield  {author} {\bibinfo {author} {\bibfnamefont {K.}~\bibnamefont {{Sun}}}\ and\ \bibinfo {author} {\bibfnamefont {H.}~\bibnamefont {{Hu}}},\ }\bibfield  {title} {\enquote {\bibinfo {title} {{Lyapunov formulation of band theory for disordered non-Hermitian systems}},}\ }\href {\doibase 10.48550/arXiv.2507.09447} {\bibfield  {journal} {\bibinfo  {journal} {arXiv e-prints}\ ,\ \bibinfo {eid} {arXiv:2507.09447}} (\bibinfo {year} {2025})},\ \Eprint {http://arxiv.org/abs/2507.09447} {arXiv:2507.09447 [cond-mat.dis-nn]} \BibitemShut {NoStop}%
\bibitem [{\citenamefont {Feinberg}\ and\ \citenamefont {Zee}(1997)}]{Feinberg1997Hermitization}%
  \BibitemOpen
  \bibfield  {author} {\bibinfo {author} {\bibfnamefont {J.}~\bibnamefont {Feinberg}}\ and\ \bibinfo {author} {\bibfnamefont {A.}~\bibnamefont {Zee}},\ }\bibfield  {title} {\enquote {\bibinfo {title} {Non-hermitian random matrix theory: Method of hermitian reduction},}\ }\href {\doibase https://doi.org/10.1016/S0550-3213(97)00502-6} {\bibfield  {journal} {\bibinfo  {journal} {Nuclear Physics B}\ }\textbf {\bibinfo {volume} {504}},\ \bibinfo {pages} {579} (\bibinfo {year} {1997})}\BibitemShut {NoStop}%
\bibitem [{\citenamefont {Claes}\ and\ \citenamefont {Hughes}(2021)}]{Claes2021SkinDisorder}%
  \BibitemOpen
  \bibfield  {author} {\bibinfo {author} {\bibfnamefont {J.}~\bibnamefont {Claes}}\ and\ \bibinfo {author} {\bibfnamefont {T.~L.}\ \bibnamefont {Hughes}},\ }\bibfield  {title} {\enquote {\bibinfo {title} {Skin effect and winding number in disordered non-hermitian systems},}\ }\href {\doibase 10.1103/PhysRevB.103.L140201} {\bibfield  {journal} {\bibinfo  {journal} {Phys. Rev. B}\ }\textbf {\bibinfo {volume} {103}},\ \bibinfo {pages} {L140201} (\bibinfo {year} {2021})}\BibitemShut {NoStop}%
\bibitem [{Sup()}]{Supp}%
  \BibitemOpen
  \href@noop {} {}\bibinfo {note} {See the Supplementary Materials at [URL will be inserted by publisher] for details of calculation.}\BibitemShut {Stop}%
\bibitem [{\citenamefont {Silvestrov}(1998)}]{Silvestrov1998localization}%
  \BibitemOpen
  \bibfield  {author} {\bibinfo {author} {\bibfnamefont {P.~G.}\ \bibnamefont {Silvestrov}},\ }\bibfield  {title} {\enquote {\bibinfo {title} {Localization in an imaginary vector potential},}\ }\href {\doibase 10.1103/PhysRevB.58.R10111} {\bibfield  {journal} {\bibinfo  {journal} {Phys. Rev. B}\ }\textbf {\bibinfo {volume} {58}},\ \bibinfo {pages} {R10111} (\bibinfo {year} {1998})}\BibitemShut {NoStop}%
\bibitem [{\citenamefont {Hatano}\ and\ \citenamefont {Nelson}(1998{\natexlab{b}})}]{Hatano1998eigenfunctions}%
  \BibitemOpen
  \bibfield  {author} {\bibinfo {author} {\bibfnamefont {N.}~\bibnamefont {Hatano}}\ and\ \bibinfo {author} {\bibfnamefont {D.~R.}\ \bibnamefont {Nelson}},\ }\bibfield  {title} {\enquote {\bibinfo {title} {Non-hermitian delocalization and eigenfunctions},}\ }\href {\doibase 10.1103/PhysRevB.58.8384} {\bibfield  {journal} {\bibinfo  {journal} {Phys. Rev. B}\ }\textbf {\bibinfo {volume} {58}},\ \bibinfo {pages} {8384} (\bibinfo {year} {1998}{\natexlab{b}})}\BibitemShut {NoStop}%
\bibitem [{\citenamefont {Longhi}(2025{\natexlab{c}})}]{Longhi2025ErraticSkin}%
  \BibitemOpen
  \bibfield  {author} {\bibinfo {author} {\bibfnamefont {S.}~\bibnamefont {Longhi}},\ }\bibfield  {title} {\enquote {\bibinfo {title} {Erratic non-hermitian skin localization},}\ }\href {\doibase 10.1103/PhysRevLett.134.196302} {\bibfield  {journal} {\bibinfo  {journal} {Phys. Rev. Lett.}\ }\textbf {\bibinfo {volume} {134}},\ \bibinfo {pages} {196302} (\bibinfo {year} {2025}{\natexlab{c}})}\BibitemShut {NoStop}%
\bibitem [{\citenamefont {Balents}\ and\ \citenamefont {Fisher}(1997)}]{Blents1997delocalization}%
  \BibitemOpen
  \bibfield  {author} {\bibinfo {author} {\bibfnamefont {L.}~\bibnamefont {Balents}}\ and\ \bibinfo {author} {\bibfnamefont {M.~P.~A.}\ \bibnamefont {Fisher}},\ }\bibfield  {title} {\enquote {\bibinfo {title} {Delocalization transition via supersymmetry in one dimension},}\ }\href {\doibase 10.1103/PhysRevB.56.12970} {\bibfield  {journal} {\bibinfo  {journal} {Phys. Rev. B}\ }\textbf {\bibinfo {volume} {56}},\ \bibinfo {pages} {12970} (\bibinfo {year} {1997})}\BibitemShut {NoStop}%
\bibitem [{\citenamefont {Lee}(2016)}]{Lee2016edgeModes}%
  \BibitemOpen
  \bibfield  {author} {\bibinfo {author} {\bibfnamefont {T.~E.}\ \bibnamefont {Lee}},\ }\bibfield  {title} {\enquote {\bibinfo {title} {Anomalous edge state in a non-hermitian lattice},}\ }\href {\doibase 10.1103/PhysRevLett.116.133903} {\bibfield  {journal} {\bibinfo  {journal} {Phys. Rev. Lett.}\ }\textbf {\bibinfo {volume} {116}},\ \bibinfo {pages} {133903} (\bibinfo {year} {2016})}\BibitemShut {NoStop}%
\bibitem [{\citenamefont {Longhi}(2020)}]{Longhi2020RandomSkin}%
  \BibitemOpen
  \bibfield  {author} {\bibinfo {author} {\bibfnamefont {S.}~\bibnamefont {Longhi}},\ }\bibfield  {title} {\enquote {\bibinfo {title} {Stochastic non-hermitian skin effect},}\ }\href {\doibase 10.1364/OL.403182} {\bibfield  {journal} {\bibinfo  {journal} {Opt. Lett.}\ }\textbf {\bibinfo {volume} {45}},\ \bibinfo {pages} {5250} (\bibinfo {year} {2020})}\BibitemShut {NoStop}%
\end{thebibliography}%


\begin{thebibliography}{4}%
\makeatletter
\providecommand \@ifxundefined [1]{%
 \@ifx{#1\undefined}
}%
\providecommand \@ifnum [1]{%
 \ifnum #1\expandafter \@firstoftwo
 \else \expandafter \@secondoftwo
 \fi
}%
\providecommand \@ifx [1]{%
 \ifx #1\expandafter \@firstoftwo
 \else \expandafter \@secondoftwo
 \fi
}%
\providecommand \natexlab [1]{#1}%
\providecommand \enquote  [1]{``#1''}%
\providecommand \bibnamefont  [1]{#1}%
\providecommand \bibfnamefont [1]{#1}%
\providecommand \citenamefont [1]{#1}%
\providecommand \href@noop [0]{\@secondoftwo}%
\providecommand \href [0]{\begingroup \@sanitize@url \@href}%
\providecommand \@href[1]{\@@startlink{#1}\@@href}%
\providecommand \@@href[1]{\endgroup#1\@@endlink}%
\providecommand \@sanitize@url [0]{\catcode `\\12\catcode `\$12\catcode
  `\&12\catcode `\#12\catcode `\^12\catcode `\_12\catcode `\%12\relax}%
\providecommand \@@startlink[1]{}%
\providecommand \@@endlink[0]{}%
\providecommand \url  [0]{\begingroup\@sanitize@url \@url }%
\providecommand \@url [1]{\endgroup\@href {#1}{\urlprefix }}%
\providecommand \urlprefix  [0]{URL }%
\providecommand \Eprint [0]{\href }%
\providecommand \doibase [0]{http://dx.doi.org/}%
\providecommand \selectlanguage [0]{\@gobble}%
\providecommand \bibinfo  [0]{\@secondoftwo}%
\providecommand \bibfield  [0]{\@secondoftwo}%
\providecommand \translation [1]{[#1]}%
\providecommand \BibitemOpen [0]{}%
\providecommand \bibitemStop [0]{}%
\providecommand \bibitemNoStop [0]{.\EOS\space}%
\providecommand \EOS [0]{\spacefactor3000\relax}%
\providecommand \BibitemShut  [1]{\csname bibitem#1\endcsname}%
\let\auto@bib@innerbib\@empty
\bibitem [{\citenamefont {Claes}\ and\ \citenamefont
  {Hughes}(2021)}]{Claes2021SkinDisorder}%
  \BibitemOpen
  \bibfield  {author} {\bibinfo {author} {\bibfnamefont {J.}~\bibnamefont
  {Claes}}\ and\ \bibinfo {author} {\bibfnamefont {T.~L.}\ \bibnamefont
  {Hughes}},\ }\bibfield  {title} {\enquote {\bibinfo {title} {Skin effect and
  winding number in disordered non-hermitian systems},}\ }\href {\doibase
  10.1103/PhysRevB.103.L140201} {\bibfield  {journal} {\bibinfo  {journal}
  {Phys. Rev. B}\ }\textbf {\bibinfo {volume} {103}},\ \bibinfo {pages}
  {L140201} (\bibinfo {year} {2021})}\BibitemShut {NoStop}%
\bibitem [{\citenamefont {Kogan}\ \emph {et~al.}(1996)\citenamefont {Kogan},
  \citenamefont {Mudry},\ and\ \citenamefont {Tsvelik}}]{Kogan1996randomDirac}%
  \BibitemOpen
  \bibfield  {author} {\bibinfo {author} {\bibfnamefont {I.~I.}\ \bibnamefont
  {Kogan}}, \bibinfo {author} {\bibfnamefont {C.}~\bibnamefont {Mudry}}, \ and\
  \bibinfo {author} {\bibfnamefont {A.~M.}\ \bibnamefont {Tsvelik}},\
  }\bibfield  {title} {\enquote {\bibinfo {title} {Liouville theory as a model
  for prelocalized states in disordered conductors},}\ }\href {\doibase
  10.1103/PhysRevLett.77.707} {\bibfield  {journal} {\bibinfo  {journal} {Phys.
  Rev. Lett.}\ }\textbf {\bibinfo {volume} {77}},\ \bibinfo {pages} {707}
  (\bibinfo {year} {1996})}\BibitemShut {NoStop}%
\bibitem [{\citenamefont {Shelton}\ and\ \citenamefont
  {Tsvelik}(1998)}]{Shelton1998LiouvilleQM}%
  \BibitemOpen
  \bibfield  {author} {\bibinfo {author} {\bibfnamefont {D.~G.}\ \bibnamefont
  {Shelton}}\ and\ \bibinfo {author} {\bibfnamefont {A.~M.}\ \bibnamefont
  {Tsvelik}},\ }\bibfield  {title} {\enquote {\bibinfo {title} {Effective
  theory for midgap states in doped spin-ladder and spin-peierls systems:
  Liouville quantum mechanics},}\ }\href {\doibase 10.1103/PhysRevB.57.14242}
  {\bibfield  {journal} {\bibinfo  {journal} {Phys. Rev. B}\ }\textbf {\bibinfo
  {volume} {57}},\ \bibinfo {pages} {14242} (\bibinfo {year}
  {1998})}\BibitemShut {NoStop}%
\bibitem [{\citenamefont {Balents}\ and\ \citenamefont
  {Fisher}(1997)}]{Blents1997delocalization}%
  \BibitemOpen
  \bibfield  {author} {\bibinfo {author} {\bibfnamefont {L.}~\bibnamefont
  {Balents}}\ and\ \bibinfo {author} {\bibfnamefont {M.~P.~A.}\ \bibnamefont
  {Fisher}},\ }\bibfield  {title} {\enquote {\bibinfo {title} {Delocalization
  transition via supersymmetry in one dimension},}\ }\href {\doibase
  10.1103/PhysRevB.56.12970} {\bibfield  {journal} {\bibinfo  {journal} {Phys.
  Rev. B}\ }\textbf {\bibinfo {volume} {56}},\ \bibinfo {pages} {12970}
  (\bibinfo {year} {1997})}\BibitemShut {NoStop}%
\end{thebibliography}%

\end{document}


\title{Supplemental Material for ``Non-Hermitian Delocalization Realizes Random Dirac Criticality in One Dimension"}

\author{Bo Li}
\email{libphysics@xjtu.edu.cn}
\affiliation{MOE Key Laboratory for Nonequilibrium Synthesis and Modulation of Condensed Matter,\\
Shaanxi Province Key Laboratory of Quantum Information and Quantum Optoelectronic Devices,\\
School of Physics, Xi’an Jiaotong University, Xi’an 710049, China}

\author{Shen Zhang}
\affiliation{MOE Key Laboratory for Nonequilibrium Synthesis and Modulation of Condensed Matter,\\
Shaanxi Province Key Laboratory of Quantum Information and Quantum Optoelectronic Devices,\\
School of Physics, Xi’an Jiaotong University, Xi’an 710049, China}

\author{Ren Zhang}
\affiliation{MOE Key Laboratory for Nonequilibrium Synthesis and Modulation of Condensed Matter,\\
Shaanxi Province Key Laboratory of Quantum Information and Quantum Optoelectronic Devices,\\
School of Physics, Xi’an Jiaotong University, Xi’an 710049, China}

\maketitle

\onecolumngrid

\tableofcontents

\section{Topological invariant}

In this section, following Ref.~\cite{Claes2021SkinDisorder}, we demonstrate that the spectral winding number of a non-Hermitian Hamiltonian $H_0$ is equivalent to the topological invariant of its Hermitized counterpart $\tilde{H}$, which belongs to class AIII in the Altland–Zirnbauer tenfold classification.\\

\textbf{Spectral winding in real space.} In the presence of disorder, the spectral winding for a non-Hermitian Hamiltonian $H_0$ is given by 
\begin{eqnarray}\label{eq:spectralwinding}
w(E)=\frac{1}{L}\text{Tr}\Big(Q^\dagger[Q,X]\Big)    
\end{eqnarray}
where $L$ is the system size, $X$ is the position operator, and $Q$ comes from polar decomposition:
\begin{eqnarray}\label{eq:polardecomposition}
E-H_0=QP.    
\end{eqnarray}
Here, $Q$ is a unitary matrix, $Q^\dagger Q=\mathbbm{1}$, and $P$ is positively defined. Without disorder, the spectral winding Eq.~\eqref{eq:spectralwinding} is reduced to the typical definition of spectral winding in momentum space. With translational symmetry, all the operators are block diagonalized in momentum space, i.e., 
\begin{eqnarray}
Q\rightarrow Q_m=\oplus_k Q_k,\qquad X\rightarrow X_m=\oplus_k(-i\partial_k),    
\end{eqnarray}
where the subscript $m$ stands for momentum space. Therefore, the winding number is rewritten as
\begin{eqnarray}
w(E)=\frac{1}{L}\text{Tr}_{m}\Big(Q_m^\dagger[Q_m,X_m]\Big)=\int_0^{2\pi}\frac{dk}{2\pi} \text{tr}\Big(Q_k^\dagger [Q_k,-i\partial_k]\Big)=\int_0^{2\pi} \frac{dk}{2\pi i}\text{tr}[Q^\dagger_k\partial_k Q_k]
=\int_0^{2\pi} \frac{dk}{2\pi i}\partial_k\ln(\det Q_k).
\end{eqnarray}
Here, $\text{Tr}(\cdots)$ acts on all freedoms, including spatial and orbital, while $\text{tr}(\cdots)$ only acts on orbital freedoms.
On the other hand, given that $E-H_0(k)=Q_kP_k$ with $Q_k^\dagger Q_k=\mathbbm 1$ and $P_k$ is positively defined, the momentum space spectral winding reads
\begin{eqnarray}
w(E)=\frac{1}{2\pi i}\int_0^{2\pi} dk\partial_k\ln\det [E-H_{0}(k)]=\frac{1}{2\pi i}\int_0^{2\pi} dk\partial_k\Big(\ln\det Q_k+\ln\det P_k\Big)=\frac{1}{2\pi i}\int_0^{2\pi} dk\partial_k(\ln\det Q_k),    
\end{eqnarray}
where $\det P_k$ is a real number such that $\int_0^{2\pi} dk\partial_k(\ln\det P_k)=\ln(\det P_{k=2\pi}/\det P_{k=0})=0$.\\

\textbf{Topological invariant in AIII class.} The Hermitized Hamiltonian
\begin{eqnarray}
\tilde{H}=\left(\begin{array}{cc}
   0  & E-H_0 \\
  E^\ast-H_0^\dagger   & 0
\end{array}\right)   
\end{eqnarray}
respects a chiral symmetry: $\{\tilde{H},S\}=0$ with $S=\sigma^z\otimes\mathbbm{1}$. Therefore, it belongs to the AIII class with topological invariant defined by the projection operator $\mathcal P$, which projects states into the subspace spanned by eigenstates of $\tilde{H}$ with negative energies. Specifically, one can define a flattened Hamiltonian 
\begin{eqnarray}
\mathcal Q=\mathbbm 1-2\mathcal P=\Big(
\begin{array}{cc}
   0  & Q_{+-} \\
   Q_{-+}  & 0
\end{array}
\Big).   
\end{eqnarray}
Subsequently, the topological invariant is defined as
\begin{eqnarray}\label{eq:windingnumber}
\nu=\frac{1}{L}\text{Tr}\{Q_{-+}[Q_{+-},X]\}   
\end{eqnarray}
which is the so-called topological winding number.\\

\textbf{Equivalence between spectral winding $w(E)$ and $\nu$.} Here, we prove that these two topological invariances are equivalent. For this purpose, we use the polar decomposition Eq.~\eqref{eq:polardecomposition} to rewrite $\tilde{H}$:
\begin{eqnarray}
\tilde{H}=\Big(
\begin{array}{cc}
   0  & QP \\
   PQ^\dagger  & 0
\end{array}
\Big).    
\end{eqnarray}
Assuming $P|\lambda_n\rangle=\lambda_n|\lambda_n\rangle$ with $\lambda_n>0$, the eigenstate of $\tilde{H}$ is given by
\begin{eqnarray}
|\Psi^{\pm}_n\rangle=\frac{1}{\sqrt{2}}\Big(
\begin{array}{cc}
    \pm Q|\lambda_n\rangle \\
     |\lambda_n\rangle 
\end{array}\Big),    
\end{eqnarray}
which satisfies $\tilde{H}|\Psi^{\pm}_n\rangle =\pm\lambda_n|\Psi_n^{\pm}\rangle$. The flatten Hamiltonian of $\Tilde{H}$ is given by 
\begin{eqnarray}
\mathcal Q=\mathbbm 1-2\sum_n|\Psi_n^{-}\rangle\langle \Psi_n^-|=\mathbbm 1-
\Big(
\begin{array}{cc}
   \mathbbm 1  & -Q \\
   -Q^\dagger  & \mathbbm 1
\end{array}
\Big)=\Big(
\begin{array}{cc}
   0  & Q \\
   Q^\dagger  & 0
\end{array}
\Big).    
\end{eqnarray}
Consequently, the components of $\mathcal Q$ are identified as $Q_{+-}=Q$ and $Q_{-+}=Q^\dagger$. Substituting these relations into Eq.~\eqref{eq:windingnumber} can convert it to Eq.~\eqref{eq:spectralwinding}, proving the equivalence of spectral winding and the winding number in AIII class.

\section{Numerical data and method}

\subsection{Transfer matrix method}

In Fig.~1(e) of the main text, we show the localization length of the Hatano–Nelson model as a function of the complex energy $E=E_r+iE_i$. For completeness, we briefly outline the transfer-matrix method used to extract the localization length from the eigenvalue equation. The eigenvalue equation for the Hatano-Nelson model is given by 
\begin{equation}
(t-\gamma)\psi_{j-1}+V_j\psi_j+(t+\gamma)\psi_{j+1}=E\psi_j,   
\end{equation}
This can be re-expressed as
\begin{equation}
\left(
\begin{array}{cc}
     \psi_{j+1}  \\
     \psi_j
\end{array}\right)=T_j
\left(
\begin{array}{cc}
     \psi_{j}  \\
     \psi_{j-1}
\end{array}\right),\qquad
T_j=\left(
\begin{array}{cc}
 \frac{E-V_j}{t+\gamma}    &-\frac{t-\gamma}{t+\gamma}  \\
   1  & 0
\end{array}
\right).
\end{equation}
Here, $T_j$ is the transfer matrix that connects the wave function at adjacent sites. 
The total transfer matrix connecting the two ends of the eigenstate is given by 
\begin{equation}
Q_L=T_LT_{L-1}\cdots T_2T_1.
\end{equation}
The localization length can be extracted from the total transfer matrix as follows
\begin{equation}
\frac{1}{\xi(E)}=\lim_{L\rightarrow 0}\frac{1}{L}\ln||Q_L||    
\end{equation}
where
\begin{equation}
||Q_L||=\text{max}_{i}\sqrt{\lambda_i},    
\end{equation}
with $\lambda_i$ being the eigenvalue of $Q_L^\dagger Q_L$. Notably, the transfer matrix is not sensitive to the boundary condition; instead, it is fully determined by the energy and disorder.\\

\begin{figure}
    \centering
    \includegraphics[width=0.5\linewidth]{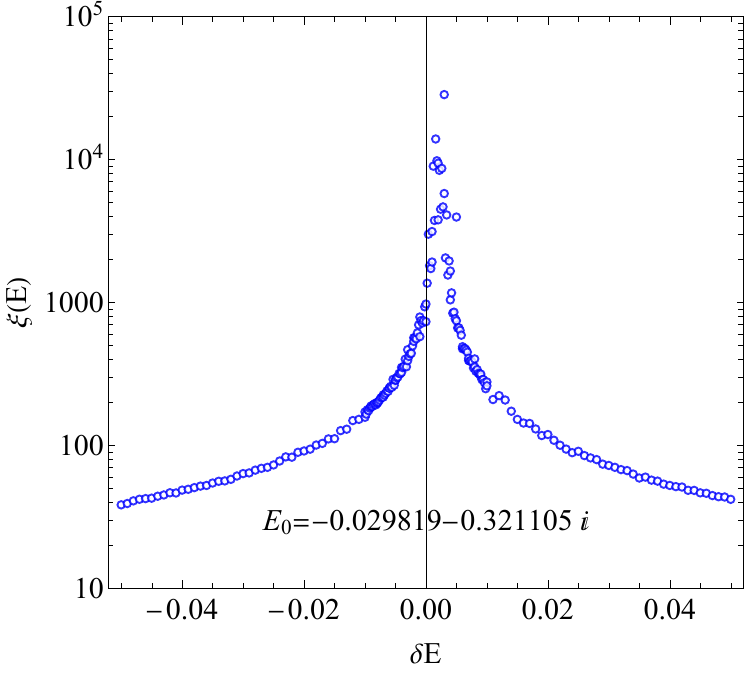}
    \caption{Localization length $\xi(E)$ with $E=E_0+ i\delta E$ as a function of $\delta E$.}
    \label{fig:diverge}
\end{figure}

In Fig.~1(e) of the main text, the localization length attains its maximum value along the PBC spectral loop. To further verify that this maximum corresponds to a divergence, we examine its variation in the vicinity of the loop. As shown in Fig.~\ref{fig:diverge}, the localization length exhibits a clear singularity when the energy crosses the spectral loop, i.e., by tuning $\delta E$ in $E=E_0+i\delta E$.

\subsection{The spatial aggregation of PBC loop states} 

\begin{figure}
    \centering
    \begin{tabular}{cc}    \includegraphics[width=0.3\linewidth]{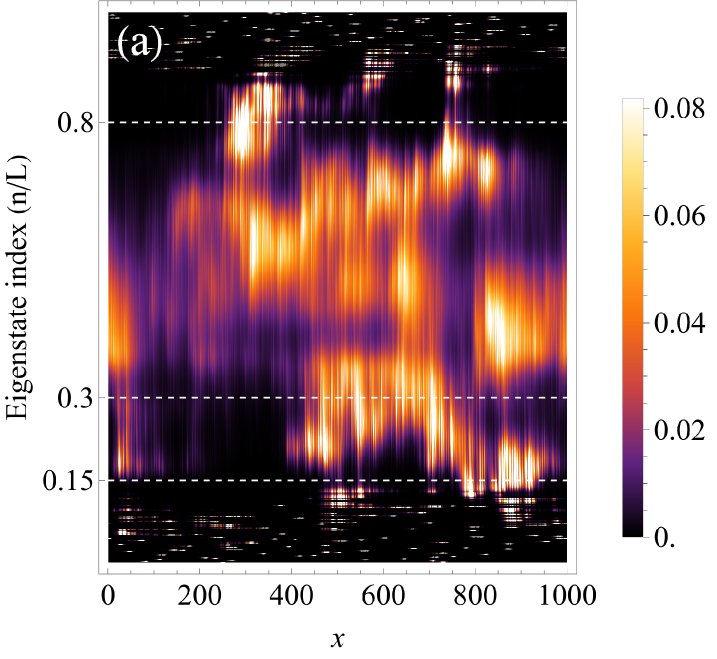}
     &  \includegraphics[width=0.6\linewidth]{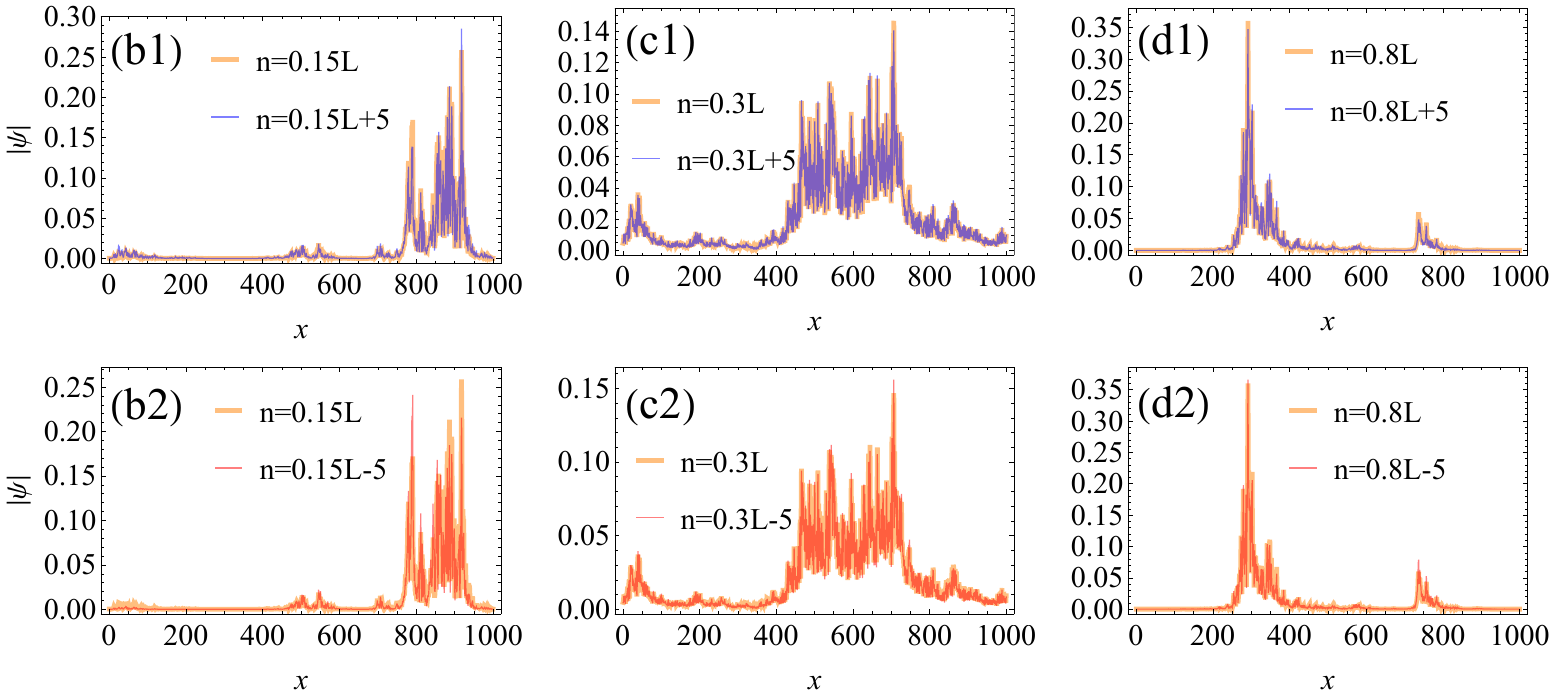} 
    \end{tabular}    
    \caption{(a) Eigenstate profiles for a specific disorder realization, with eigenstates arranged in ascending order according to the real part of their eigenvalues. (b1–d1) and (b2–d2) demonstrate that eigenstates adjacent in this ordering exhibit similar spatial distributions.}
    \label{fig:wave_similar}
\end{figure}

In the main text, Fig.~2(c) shows that the eigenstate distribution depends strongly on the disorder configuration and is highly non-uniform. A long-wavelength analysis suggests that loop-state profiles tend to cluster around a few spatial positions, as illustrated in Fig.~\ref{fig:wave_similar}(a). Figures~\ref{fig:wave_similar}(b1–d1) and (b2–d2) further demonstrate that eigenstates with nearly identical real parts of the eigenvalues exhibit highly similar spatial distributions. This observation supports our long-wavelength analysis, where such eigenstates are selected across different disorder configurations.

\section{Wave function correlation}

In this section, we elaborate on the calculation details for the wavefunction correlation at the topological Anderson transition. For completeness, we first recap the building of the random Dirac Hamiltonian via the Hermitization method and the construction of the zero-mode ansatz, then give the details of calculating the eigenstate-profile correlation by making a hard-wall approximation. For $W=0$, the momentum space Hamiltonian is expressed as 
\begin{equation}\label{eq:Hermitized}
\Tilde{H}=\left(
\begin{array}{cc}
   0  & E-2t\cos k-i2\gamma\sin k \\
  E^\ast-2t\cos k+i2\gamma\sin k   & 0
\end{array}
\right)    
\end{equation}
which is gapless at $k_0$ that satisfies $E_r=2t\cos k_0$ and $E_i=2\gamma\sin k_0$. The long-wavelength Hamiltonian around $k_0$ reads
\begin{equation}
\Tilde{H}(k_0+q)=(\frac{tE_i}{\gamma}\sigma^x+\frac{\gamma E_r}{t}\sigma^y)q+O(q^2).
\end{equation}
In the long-wavelength limit, adding the effect of disorder and turning to real space, we end up with  the following Hamiltonian
\begin{equation}
\Tilde{H}_{eff}=(\eta_x\sigma^x+\eta_y\sigma^y)(-i\partial_x)+m(x)\sigma^x   
\end{equation}
where $\eta_x=t E_i/\gamma$, $\eta_y=\gamma E_r/t$, and $m(x)$ originates from the long-wavelength component of the on-site potential $V_j$, satisfying 
\begin{equation}\label{eq:noise}
\langle m(x) m(x^\prime)\rangle_{dis}=g\delta(x-x^\prime).    
\end{equation}
Here, we should be aware that the disorder field $m(x)$ depends on the position of the complex energy on the loop because it should match the spatial pattern of the corresponding long-wavelength expansion of the Hamiltonian.\\

We assume a zero-energy state ansatz
\begin{equation}\label{eq:ansatz}
\Psi_0(x)=\frac{1}{\sqrt{\mathcal N}}\Vec{\chi} \exp [\alpha V(x)],\qquad    V(x)=\int_0^x dym(y),    
\end{equation}
where $\Vec{\chi}$ is a spinor in the sublattice space, $\alpha=\alpha_r+i\alpha_i$, and $\mathcal N$ is the normalization factor taking the form
\begin{equation}
\mathcal N=\int_{-L/2}^{L/2} dx\exp [2\alpha_rV(x)].    
\end{equation}
Substituting the ansatz into the eigen equation, zero energy requires
\begin{equation}
[(1-i\alpha\eta_x)\sigma^x-i\alpha\eta_y\sigma^y]\Vec{\chi}=0.  
\end{equation}
This condition reduces to
\begin{eqnarray}
\det [(1-i\alpha\eta_x)\sigma^x-i\alpha\eta_y\sigma^y]=0,   
\end{eqnarray}
which is solved by
\begin{eqnarray}\label{eq:anstz_coefficient}
\alpha_1=\frac{\eta_y-i\eta_x}{\eta_x^2+\eta^2_y},\qquad\text{or}\qquad
\alpha_2=\frac{-\eta_y-i\eta_x}{\eta_x^2+\eta^2_y}.
\end{eqnarray}
As discussed in the main text, we are allowed to set $\alpha_r=1$ for convenience. By following the Kogan-Mudry-Tsvelik method~\cite{Kogan1996randomDirac, Shelton1998LiouvilleQM}, the variable $V(x)$ in Eq.~\eqref{eq:ansatz} behaves like the position variable in a random walk where $x$ is regarded as the ``time". The variable $V(x)$ respect the following Gaussian distribution  
\begin{equation}
P(V)DV=\exp\Big[-\frac{1}{2g}\int dx(\partial_xV)^2\Big].    
\end{equation}
Our goal is to calculate the spatial correlation of the zero-mode profile:
\begin{equation}
W_q(x,L)=\frac{1}{Z_0}\int DV \Big|\frac{1}{\sqrt{\mathcal N}}e^{V(x)}\frac{1}{\sqrt{\mathcal N}}e^{V(0)}\Big|^q e^{-S_0}    
\end{equation}
where $S_0=\frac{1}{2g}\int dx(\partial_xV)^2$, and $Z_0=\int DV e^{-S_0}$. In the calculation, we will take a periodic boundary condition $V(-L/2)=V(L/2)$. The normalization factor can be calculated by the following relation
\begin{eqnarray}
\mathcal N^{-q}=\frac{1}{\Gamma(q)}\int_0^\infty d\omega \omega^{q-1}e^{-\omega\mathcal N},   
\end{eqnarray}
where $\Gamma(q)$ is the Gamma function. Therefore, 
\begin{eqnarray}\label{eq:correlation}
 W_q(x,L)=\frac{1}{\Gamma(q)}\int_0^\infty d\omega \omega^{q-1}\langle e^{qV(x)}e^{qV(0)}\rangle_{dis}   
\end{eqnarray}
with
\begin{eqnarray}\label{eq:average}
\langle e^{qV(x)}e^{qV(0)}\rangle_{dis}=\frac{1}{Z_0}\int DV e^{qV(x)}e^{qV(0)}e^{-S}   
\end{eqnarray}
where 
\begin{eqnarray}
S=\int_0^L dx\Big[ \frac{1}{2g}(\partial_x V)^2+\omega e^{2V}\Big].    
\end{eqnarray}
In the next step, we follow Ref.~\cite{Blents1997delocalization} to obtain the correlation. Eq.~\eqref{eq:average} describes a system that evolves according to the Liouville Hamiltonian
\begin{eqnarray}
H=-\frac{g}{2}\frac{d^2}{dV^2}+\omega e^{2V}.    
\end{eqnarray}
One can work out Eq.~\eqref{eq:average} in the operator presentation
\begin{eqnarray}\label{eq:correlator}
\langle e^{qV(x)}e^{qV(0)}\rangle_{dis}=\frac{\langle 0|e^{-LH/2}e^{qV(x)}e^{qV(0)}e^{-LH/2}|0\rangle}{\langle 0|e^{-LH_0}|0\rangle}    
\end{eqnarray}
where $e^{qV(x)}=e^{xH}e^{qV(0)}e^{-xH}$, and $H_0=-\frac{g}{2}\frac{d^2}{dV^2}$.  The denominator represents that a free particle spreads in space, and it can be shown that
\begin{eqnarray}\label{eq:denominator}
\langle 0|e^{-LH_0}|0\rangle= \frac{1}{\sqrt{4\pi L}}.   
\end{eqnarray}
To calculate the numerator, a simple way is to make a hard-wall approximation, namely, replacing the exponential potential by the following potential:
\begin{eqnarray}
U(V)=
\begin{cases}
0, & V< V_\omega  \\
\infty, & \text{Otherwise}  
\end{cases}.
\end{eqnarray}
Here, the wall position is $V_\omega= \frac{1}{2}\ln\frac{1}{\omega}$, which is identified by the condition $\omega e^{2V_\omega}=1$. In the approximated potential, the random walk position $V(x)$ is confined to $V<V_\omega$. For the convenience of calculation, one can redefine the coordinates by making the replacement $V\rightarrow V_\omega- V$, such that the domain of $V$ is changed to $V>0$. In the new frame, the numerator becomes
\begin{eqnarray}\label{eq:numerator}
\langle 0|e^{-LH/2}e^{qV(x)}e^{qV(0)}e^{-LH/2}|0\rangle&&\rightarrow
\langle V_\omega|e^{-LH_0/2}e^{q(V_\omega-V(x))}e^{q(V_\omega-V(0))}e^{-LH_0/2}|V_\omega\rangle\nonumber\\
&&=e^{2qV_\omega}\langle V_\omega|e^{-LH_0/2}e^{-qV(x)}e^{-qV(0)}e^{-LH_0/2}|V_\omega\rangle.
\end{eqnarray}
Now, plugging Eqs.~\eqref{eq:numerator} and~\eqref{eq:denominator} into Eqs.~\eqref{eq:correlator} and~\eqref{eq:correlation}, one obtains
\begin{eqnarray}\label{eq:correlation1}
W_q(x,L)=\frac{\sqrt{4\pi L}}{\Gamma(q)}\int_0^\infty\frac{d\omega}{\omega}\langle V_\omega|e^{-LH_0/2}e^{-qV(x)}e^{-qV(0)}e^{-LH_0/2}|V_\omega\rangle   
\end{eqnarray}
where we used $e^{2qV_\omega}=\omega^{-q}$. To proceed, one can introduce a set of standing wave bases $|k\rangle$ that satisfy
\begin{eqnarray}
&&\langle V|k\rangle=\sqrt{\frac{2}{\pi}}\sin(kV), \nonumber\\
&&\langle k| k^\prime\rangle =\delta(k-k^\prime),\nonumber\\
&&\int_0^\infty dk |k\rangle\langle k|=1.
\end{eqnarray}
By inserting these bases, the evolution is presented as 
\begin{eqnarray}\label{eq:evolution}
&&\langle V_\omega|e^{-LH_0/2}e^{-qV(x)}e^{-qV(0)}e^{-LH_0/2}|V_\omega\rangle\nonumber\\
=&&\langle V_\omega|e^{-LH_0/2}e^{xH_0}e^{-qV(0)}e^{-xH_0}e^{-qV(0)}e^{-LH_0/2}|V_\omega\rangle\nonumber\\
=&&\int dk_1 dk_2 dk_3 \langle V_\omega|e^{-LH_0/2}e^{xH_0}|k_1\rangle\langle k_1|e^{-qV(0)}e^{-xH_0}|k_2\rangle\langle k_2|e^{-qV(0)}|k_3\rangle\langle k_3|e^{-LH_0/2}|V_\omega\rangle\nonumber\\
=&&\int dk_1 dk_2 dk_3 e^{-(\frac{L}{2}-x)k_1^2}\langle V_\omega|k_1\rangle\langle k_1|e^{-qV(0)}|k_2\rangle e^{-xk_2^2}\langle k_2|e^{-qV(0)}|k_3\rangle\langle k_3|V_\omega\rangle e^{-\frac{L}{2}k_3^2},
\end{eqnarray}
where we used that $H_0|k\rangle= k^2|k\rangle$. Note the element
\begin{eqnarray}
\langle k|e^{-qV}|k^\prime\rangle=\int_0^\infty dV 
\frac{2}{\pi}\sin(k V)\sin(k^\prime V)e^{-qV}=\frac{4qkk^\prime}{\pi[q^2+(k-k^\prime)^2][q^2+(k+k^\prime)^2]}.
\end{eqnarray}
Using this relation, the last line in Eq.~\eqref{eq:evolution} becomes
\begin{eqnarray}
&&\int_0^\infty dk_1 dk_2 dk_3 e^{-(\frac{L}{2}-x)k_1^2}e^{-xk_2^2}e^{-\frac{L}{2}k_3^2}\frac{2}{\pi}\sin(k_1V_\omega)\sin(k_3V_\omega)\nonumber\\
&&\qquad\qquad\quad\times\frac{4qk_1k_2}{\pi[q^2+(k_1-k_2)^2][q^2+(k_1+k_2)^2]} 
\frac{4qk_2k_3}{\pi[q^2+(k_2-k_3)^2][q^2+(k_2+k_3)^2]} \nonumber\\
\approx&&\frac{16q^2}{\pi^2}\int_0^\infty dk_2\frac{k_2^2e^{-xk_2^2}}{(q^2+k_2^2)^4}
\Big[\sqrt{\frac{2}{\pi}}\int dk_1 e^{-(\frac{L}{2}-x)k_1^2}k_1\sin(k_1 V_\omega)\Big] \Big[\sqrt{\frac{2}{\pi}}\int dk_3 e^{-\frac{L}{2}k_3^2}k_3\sin(k_3 V_\omega)\Big]\nonumber\\
=&&\frac{16q^2}{\pi^2}\int_0^\infty dk_2\frac{k_2^2e^{-xk_2^2}}{(q^2+k_2^2)^4}\Big[\frac{V_\omega}{L^{3/2}}e^{-V_\omega^2/(2L)}\Big]\Big[\frac{V_\omega}{(L-2x)^{3/2}}e^{-V_\omega^2/(2L-4x)}\Big]\nonumber\\
\approx&&\frac{16q^2}{\pi^2}\frac{V_\omega^2}{L^3}e^{-V_\omega^2/L}\int_0^\infty dk\frac{k^2e^{-xk^2}}{(q^2+k^2)^4}.
\end{eqnarray}
In the third line above, we used that, in the limit $L\gg 1$, the integral is dominated by $k_1, k_3\approx 0$; in the last step, we assumed $L\gg x$ and replaced $k_2$ by $k$.
Substituting this result into  Eq.~\eqref{eq:correlation1} and employing the relation $d\omega/\omega=-2dV_\omega$, one obtains
\begin{eqnarray}\label{eq:correlation2}
W_q(x,L)=\frac{\sqrt{4\pi L}}{\Gamma(q)}\int_{-\infty}^\infty 2dV_\omega \frac{16q^2}{\pi^2}\frac{V_\omega^2}{L^3}e^{-V_\omega^2/L}I(x,q)=\frac{32 q^2}{\Gamma(q)\pi L}I(x,q),
\end{eqnarray}
where 
\begin{eqnarray}\label{eq:correlation}
I(x,q)&&=\int_0^\infty dk\frac{k^2e^{-xk^2}}{(q^2+k^2)^4} =\frac{1}{q^5} \int_0^\infty dk\frac{k^2e^{-\Tilde{x}k^2}}{(1+k^2)^4}\nonumber\\ 
&&=\frac{1}{96 q^5}\Big[
-2\sqrt{\pi\Tilde{x}}(-3+4\Tilde{x}(1+\Tilde{x}))+\pi e^{\Tilde{x}}\Big(3+2\title{x}(-3+6\Tilde{x}+4\Tilde{x}^2)erf(\sqrt{\Tilde{x}})\Big)
\Big]
\end{eqnarray}
where $\Tilde{x}=xq^2$.\\

Note that in the limit $x\gg 1$, the exponential factor $e^{-xk^2}$ decays quickly, so only $0<k\lesssim x^{-1/2}\ll 1$ contribute to the integral significantly. As a result,
\begin{eqnarray}
I(x,q)\approx\int_0^\infty dk k^2 e^{-xk^2}\Big(1-\frac{4k^2}{q^2}\Big)+\cdots=\frac{\sqrt{\pi}}{4q^8 x^{3/2}}-\frac{3\sqrt{\pi}}{2q^{10} x^{5/2}}+\cdots\sim x^{-3/2}.   
\end{eqnarray}
Substituting this result into Eq.~\eqref{eq:correlation2}, one finally obtains the scaling of the wavefunction correlation
\begin{eqnarray}
 W_q(x,L)\sim L^{-1}x^{-3/2},   
\end{eqnarray}
which is entirely independent of $q$. Above, the $L^{-1}$ is a characteristic of localization; however, the power-law decay $x^{-3/2}$ is a manifestation of criticality. As a comparison, an exponentially localized wavefunction exhibits exponential decay in wavefunction correlation.


\begin{figure}
    \centering
    \begin{tabular}{ccc} \includegraphics[width=0.3\linewidth]{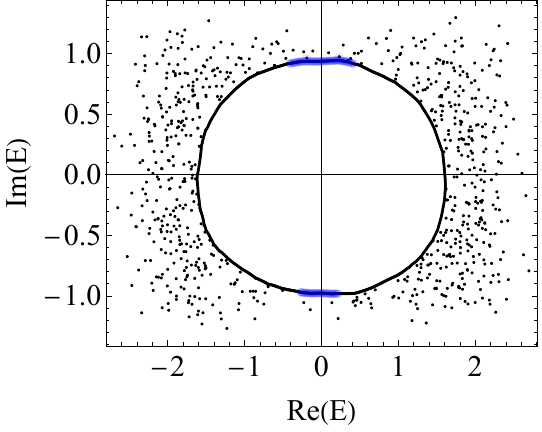}  & 
    \includegraphics[width=0.3\linewidth]{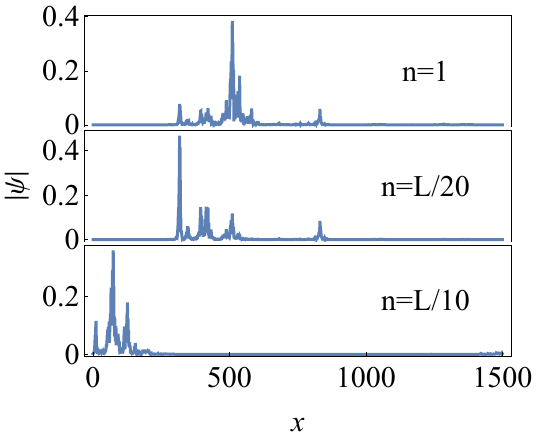} &  \includegraphics[width=0.3\linewidth]{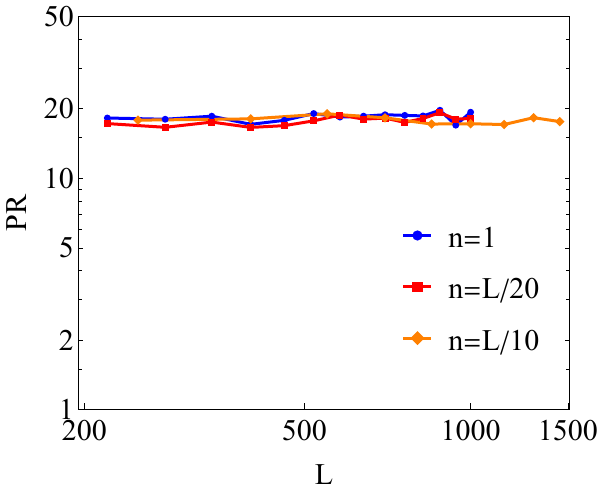}
    \end{tabular}
    \caption{(a) The spectrum of the Hatano-Nelson model with a complex on-site disorder. Here, the blue segment shows the first $1/10$ eigenvalues, with all eigenvalues ordered by their absolute values in descending order. (b) The three wavefunction profiles $|\psi_n|$ for $n=1, L/20,L/10$. (c) The participation ratio for the $n$-th eigenstate as a function of system size. Parameters are $t=1$, $\gamma=0.5$, and $W=1.5$.}
    \label{fig:complex_HN}
\end{figure}

\section{Other models---long-wavelength behavior}

\subsection{Hatano-Nelson model with complex-valued disorder}


We consider the modified Hatano-Nelson model where the disorder takes complex values, i.e., $V_j=u_j+iv_j$ with $u_j\in[-W_1, W_1]$ and $v_j\in[-W_2, W_2]$. Performing similar numerical analysis, we find that, for the delocalized modes, their participation ratios also approach a finite value as the system size grows to a sufficiently large value, see Fig.~\ref{fig:complex_HN}. Following the previous method, we can also construct the zero-mode wavefunction in this case. As before, the long-wavelength Hamiltonian is given by
\begin{eqnarray}\label{eq:randomDirac}
\Tilde{H}_{eff}=(\eta_x\sigma^x+\eta_y\sigma^y)(-i\partial_x)+u(x)\sigma^x +v(x)\sigma^y=\sigma^x \mathcal M(x),    
\end{eqnarray}
where 
\begin{eqnarray}
\mathcal M(x)= (\eta_x+i\eta_y\sigma^z)(-i\partial_x)+u(x) +v(x)\sigma^z.    
\end{eqnarray}
Based on the form of the Hamiltonian, the zero-mode ansatz is assumed to be 
\begin{eqnarray}
\Psi_0(x)=\mathcal N^{-1/2}e^{\alpha\int_0^x dy[u(y)+iv(y)\sigma^z]}\Vec{\chi},  
\end{eqnarray}
where $\mathcal N$ is a normalization factor. Acting Hamiltonian on this trial function, it requires
\begin{eqnarray}
\sigma^x\Big[u(x)+iv(x)\sigma^z\Big]\Big[1-i\alpha(\eta_x+i\eta_y\sigma^z)\Big]\Vec{\chi}=0,    
\end{eqnarray}
which is reduced to $\det [1-i\alpha(\eta_x+i\eta_y\sigma^z)]=0$. This results in 
\begin{eqnarray}
\alpha_1=\frac{-\eta_y+i\eta_x}{\eta_x^2+\eta^2_y},\qquad\text{or}\qquad
\alpha_2=\frac{\eta_y+i\eta_x}{\eta_x^2+\eta^2_y},
\end{eqnarray}
which coincides with the previous case Eq.~\eqref{eq:anstz_coefficient}. According to the analysis above, it is obvious that 
\begin{eqnarray}
\Vec{\chi}_1=\left(
\begin{array}{cc}
     1  \\
     0 
\end{array}
\right)\qquad\text{or}\qquad
\Vec{\chi}_2=\left(
\begin{array}{cc}
     0  \\
     1 
\end{array}
\right). 
\end{eqnarray}
Consequently, the calculation of the correlation function resembles the process before. For instance, let us take $\Vec{\chi}_1$ as an example. The zero-mode wavefunction reads
\begin{eqnarray}
\Psi_0(x)=\mathcal N_1^{-1/2}e^ {\alpha_1\int_0^x dy[u(y)+iv(y)]}\left(
\begin{array}{cc}
     1  \\
     0 
\end{array}
\right)   
\end{eqnarray}
where $\mathcal N_1=\exp[2\int_0^x dy\Big(\alpha_{1r}u(y)-\alpha_{1i} v(y)\Big)]$, with $\alpha_{1r}, \alpha_{1i}$ being the real and imaginary part of $\alpha_1$, respectively. Given that $|\Psi_0(x)|\propto \exp[\int_0^x dy\Big(\alpha_{1r}u(y)-\alpha_{1i} v(y)\Big)]$, the current scenario is the converted to the previous case if we let $m(x)=\alpha_{1r}u(x)-\alpha_{1i} v(x)$.\\

\subsection{Two-sublattice model}

We consider the two-band Hamiltonian with NHSE in the main text: 
\begin{eqnarray}
H_0= (v+r\cos k)\tau^x+(r\sin k+i\frac{\lambda}{2})\tau^z.    
\end{eqnarray}
For a given complex energy $E\in\mathbbm C$, the corresponding Hermitized Hamiltonian $\Tilde{H}$ is gapless at 
$k_0$ satisfying $E^2=(v+r\cos k_0)^2+(r\sin k_0+i\frac{\lambda}{2})^2$. At the momentum $k_0$, the Hamiltonian $\Tilde{H}(k_0)$ has a null space spanned by
\begin{eqnarray}
|1\rangle=\frac{1}{\sqrt{\Omega}}\left(
\begin{array}{cc}
     0  \\
     1 
\end{array}
\right)\otimes|\varphi\rangle\qquad
|2\rangle=\frac{1}{\sqrt{\Omega}}\left(
\begin{array}{cc}
     1 \\
     0 
\end{array}
\right)\otimes|\Tilde{\varphi}\rangle 
\end{eqnarray}
where
\begin{eqnarray}
|\varphi\rangle=(b+\sqrt{a^2+b^2}, a)^T,\qquad
|\Tilde{\varphi}\rangle=(b^\ast+\sqrt{a^2+b^{\ast 2}}, a)^T, 
\end{eqnarray}
and $\Omega=\langle\varphi|\varphi\rangle=\langle\Tilde{\varphi}|\Tilde{\varphi}\rangle$. Here, $a=v+r\cos k_0$ and $b=(r\sin k_0+i\frac{\lambda}{2})$. In the subspace spanned by $(|1\rangle, |2\rangle)$, the Hamiltonian $\Tilde{H}(k_0+q)$ is expanded to the linear order of momentum $q$ as
\begin{eqnarray}
 H_{eff}
 &&=q\left(
 \begin{array}{cc}
  \langle 1|H^{(1)}|1\rangle & \langle 1|H^{(1)}|2\rangle \\
 \langle 2|H^{(1)}|1\rangle  & \langle 2|H^{(1)}|2\rangle
 \end{array}
 \right)
\end{eqnarray}
where
\begin{eqnarray}
 H^{(1)}=\partial_k\Tilde{H}(k)|_{k=k_0} 
=\left(
\begin{array}{cc}
  0  & A^{(1)} \\
  A^{(1)}   & 0 
\end{array}
\right)
\qquad \text{with}\qquad
A^{(1)}=-\partial_kH_0(k)|_{k=k_0}=r(\sin k_0\tau^x-\cos k_0\tau^z).
\end{eqnarray}
It is readily checked that
\begin{eqnarray}
H_{eff}=\left(
\begin{array}{cc}
    0 & \mathcal K \\
 \mathcal K^\ast& 0
\end{array}
\right) q
\end{eqnarray}
where
\begin{eqnarray}
\mathcal K=\mathcal K_x+i\mathcal K_y=\frac{1}{\Omega}\langle\varphi | A^{(1)}|\Tilde{\varphi}\rangle=\frac{-2r(b^\ast+\sqrt{a^2+b^{\ast 2}})(b^\ast\cos k_0-a\sin k_0)}{a^2+(b+\sqrt{a^2+b^2})(b^\ast+\sqrt{a^2+b^{\ast 2}})}.
\end{eqnarray}
Therefore, by replacing $q\rightarrow-i\partial_x$, the full long-wavelength expansion of the Hamiltonian is given by
\begin{eqnarray}
\mathcal H_{eff}=(\mathcal K_x\sigma^x+\mathcal K_y\sigma^y)(-i\partial_x),   
\end{eqnarray}
where $\sigma^x, \sigma^y$ are Pauli matrices in chiral space.\\

Next, we consider the effect of a weak disorder:
\begin{eqnarray}\label{eq:disorder}
\hat H_{dis}=\sum_{j=1}^L\psi_j^\dagger\sigma^x\otimes\mathcal V_j \psi_j \qquad \text{with}\qquad 
\mathcal V_j=\left(
\begin{array}{cc}
    v_{1j} & 0 \\
    0 & v_{2j}
\end{array}
\right),
\end{eqnarray}
where $v_{1j}, v_{2j}$ are site-dependent random variables following the same  boxed distribution $v_{1j}, v_{2j}\in[-W,W]$. Here, for the convenience of notation, we relabel the independent random potential on each atom site by distinguishing their sublattice freedom. To focus on the physics near the gapless points, we expand the operator $\psi_j$ as follows:
\begin{eqnarray}\label{eq:longwave}
 \psi_j\simeq\frac{1}{\sqrt{L}}\sum_{q\ll\Lambda^{-1}}e^{i(k_0+q)j}\phi(k_0+q)
\end{eqnarray}
where $\Lambda^{-1}$ is a momentum space cut-off, and 
\begin{eqnarray}\label{eq:projection}
\phi(k_0+q)=c_{1}(q)|1\rangle+c_{2}(q)|2\rangle.    
\end{eqnarray}
Substituting Eqs.~\eqref{eq:longwave} and~\eqref{eq:projection} into Eq.~\eqref{eq:disorder} leads to 
\begin{eqnarray}
\hat{H}_{dis}&&=\frac{1}{L}\sum_{q,q^\prime}\sum_je^{-i(q-q^\prime)j}\phi^\dagger(q)\sigma^x\otimes\mathcal V_j\phi(q^\prime)\nonumber\\
&&=\frac{1}{\sqrt{L}}\sum_{q,q^\prime}\phi^\dagger(q)\sigma^x\otimes\mathcal V(q-q^\prime)\phi(q^\prime)\nonumber\\
&&=\frac{1}{\sqrt{L}}\sum_{q,q^\prime}(c^\dagger_{1}(q),c^\dagger_{2}(q))M(q-q^\prime)
\left(
\begin{array}{cc}
    c_{1}(q^\prime)  \\
    c_{2}(q^\prime) 
\end{array}
\right)
\end{eqnarray}
where $\mathcal V(q-q^\prime)=\frac{1}{\sqrt{L}}\sum_je^{-i(q-q^\prime)j}\mathcal V_j$ and 
\begin{eqnarray}
M(q-q^\prime)&&=\left(
\begin{array}{cc}
 \langle 1|\sigma^x\otimes\mathcal V(q-q^\prime)|1\rangle & \langle 1|\sigma^x\otimes\mathcal V(q-q^\prime)|2\rangle \\
\langle 2|\sigma^x\otimes\mathcal V(q-q^\prime)|1\rangle  &  \langle 2|\sigma^x\otimes\mathcal V(q-q^\prime)|2\rangle
\end{array}
\right)\nonumber\\
&&=\left(
\begin{array}{cc}
   0  & \mu^\ast v_1(q-q^\prime)+\nu v_2(q-q^\prime) \\
 \mu v_1(q-q^\prime)+\nu v_2(q-q^\prime) & 0
\end{array}
\right)\nonumber\\
&&=[\mu_r v_1(q-q^\prime)+\nu v_2(q-q^\prime)]\sigma^x-\mu_i u_1(q-q^\prime)\sigma^y.
\end{eqnarray}
Here, $\mu=(b+\sqrt{a^2+b^2})^2/\Omega$ and $\nu=a^2$; $\mu=\mu_r+i\mu_i$ (with $\mu_r, \mu_i\in\mathbbm R$) takes a complex value and $\nu$ takes a real values.  Now, we transform the long-wavelength Hamiltonian to real space by considering $v_{1,2}(x)=\frac{1}{\sqrt{L}}\sum_{q<\Lambda^{-1}} e^{iqx}v_{1,2}(q)$, the effective contribution of disorder can be expressed as
\begin{eqnarray}
\hat H_{dis}=\int dx\psi^\dagger(x)\Big([\mu_r v_1(x)+\nu v_2(x)]\sigma^x-\mu_i v_2(x)\sigma^y\Big)\psi(x),
\end{eqnarray}
where $\psi(x)=(c_{1}(x),c_{2}(x))^T$. Collecting all long-wavelength terms together, we end up with
\begin{eqnarray}\label{eq:randomDirac1}
\Tilde{H}_{eff}= (\mathcal K_x\sigma^x+\mathcal K_y\sigma^y)(-i\partial_x) +u(x)\sigma^x+v(x)\sigma^y,   
\end{eqnarray}
where $u(x)=\mu_r v_1(x)+\nu v_2(x)$ and $v(x)=-\mu_i v_2(x)$. Eq.~\eqref{eq:randomDirac1} takes the same form as Eq.~\eqref{eq:randomDirac}, so their zero-mode profiles have the same properties.

\bibliographystyle{apsrev4-1-title}
\bibliography{delocalization}